\documentclass[12pt,letterpaper]{article}

\usepackage{setspace}
\usepackage{geometry}
\usepackage[utf8]{inputenc}
\usepackage{lscape}
\usepackage{amsmath}
\usepackage{amsfonts}
\usepackage{amssymb}

\usepackage{newtxtext,newtxmath}
\usepackage{graphicx}
\usepackage[labelfont=bf, font=small]{caption}
\usepackage{subcaption}
\usepackage{booktabs}

\usepackage{float}
\restylefloat{table}

\usepackage{fancyhdr}
\usepackage{hyperref}
\usepackage{booktabs}
\usepackage{siunitx}
\usepackage[toc,page]{appendix}
\usepackage{lineno}
\usepackage{authblk}
\usepackage{bbding}
\usepackage{pseudocode}
\usepackage{mwe}
\usepackage{multirow}
\usepackage{mathtools}
\usepackage{graphicx,array}
\usepackage{algpseudocode}
\usepackage[dvipsnames]{xcolor}
\usepackage[none]{hyphenat}
\usepackage{url}
\usepackage{natbib}
\usepackage{multicol}

\author[1,2,3]{\fontsize{10}{10}\selectfont\boldmath\bfseries{Vanessa Echeverri}
}
\author[2, \Envelope]{\fontsize{10}{10}\selectfont\boldmath\bfseries{Juan C. Duque}
}

\author[2,4]{\fontsize{10}{10}\selectfont\boldmath\bfseries{Daniel E. Restrepo}
}

\affil[1]{
 Department of Mathematical Sciences. Universidad de Medell\'in. Medell\'in, Colombia.
}
\affil[2]{
 Research in Spatial Economics (RiSE) Group. Universidad EAFIT. Colombia.
}

\affil[3]{
 Department of Geography. University of California, Santa Barbara. CAL, USA.
}

\affil[4]{
 Department of Mathematics. University of Texas at Austin. TX, USA.
}

 \affil[\Envelope]{Correspondence: jduquec1@eafit.edu.co. Carrera 49, 7 Sur-50. Department of Mathematical Sciences. Universidad EAFIT. 050022, Medellín, Antioquia, Colombia.
 }

\title{\LARGE Identifying poverty traps based on the network structure of economic output}

\begin{document}
\maketitle

\begin{abstract}

In this work, we explore the relationship between monetary poverty and production combining relatedness theory, graph theory, and regression analysis. We develop two measures at product level that capture short-run and long-run patterns of poverty, respectively. We use the network of related products (or product space) and both metrics to estimate the influence of the productive structure of a country in its current and future levels of poverty. We found that poverty is highly associated with poorly connected nodes in the PS, especially products based on natural resources. We perform a series of regressions with several controls (including human capital, institutions, income, and population) to show the robustness of our measures as predictors of poverty. Finally, by means of some illustrative examples, we show how our measures distinguishes between nuanced cases of countries with similar poverty and production and identify possibilities of improving their current poverty levels.

\end{abstract}

\vspace{6pt}\noindent{\fontsize{9}{9}\selectfont\textbf{Keywords:} poverty, poverty traps, quiescence traps, product space, economic development, relatedness theory}

\vspace{6pt}\noindent{\fontsize{9}{9}\selectfont\textbf{Funding:} this article was completed with support  from the PEAK Urban programme, supported by UKRIs Global Challenge Research Fund, Grant Ref.: ES/P011055/1}

\vspace{6pt}\noindent{\fontsize{9}{9}\selectfont\textbf{Acknowledgements:} we are grateful with Professor Carlos Piedrahita from Universidad de Medell\'in for his support and suggestions in this process.}

\newpage

\begin{center}
{\fontsize{18}{18}\selectfont\boldmath\bfseries{Identifying poverty traps based on the network structure of economic output.}} 
\end{center}

\section{Introduction}

Poverty is a multidimensional and structural phenomenon. According to the United Nations (UN), poverty goes beyond the lack of resources or income, involving the incapability of people to fulfill their basic needs or ensure sustainable livelihoods  \citep{unendingpov}. Poverty is manifested as freedom privation in several ways, such as in economic, political, and social scenarios, going from starvation to limited participation in society's decision-making \citep{sen2014development}. Poverty reduction has been a central target of public policy during the last century, and great goals have been accomplished in this direction. According to the World Bank \citep{world2018decline},  the number  of people living in extreme poverty in the world fell to 736 million people, which amounts to a reduction of more than a billion people living in poverty over the last 25 years. Despite this, the poverty reduction rates decelerated in 2015 falling below values of 3\%, in contrast with the average rate of the last 25 years, which reduced overall, from 36\% to 10\% depending on the country \citep{world2018decline}.\\

Partly, this deceleration of poverty reduction is associated with a fraction of the world population that remains in poverty regardless of the efforts of international or national institutions. These people or societies live in a situation that encompasses several complex phenomena that reinforce each other and reduce the possibilities of any improvement in quality of life, e.g., lack of education, insufficient civil or property rights, corruption, and weak and unstable institutions. This situation of everlasting poverty or the incapability of breaking the reinforcing poverty cycle is called poverty trap \citep{azariadis2005poverty}.\\

The quantitative approaches to model poverty traps have followed the lines of the seminal paper of \cite{solow1956contribution}, i.e., economic growth models with multiple equilibria, where some of them correspond to low levels of human or physical capital accumulation. From this perspective, an economy falls into a (neoclassical) poverty trap if it attains a stable equilibrium point associated with low-income levels (\citealp{barro2004economic};     \citealp{azariadis2005poverty}; \citealp{kraay2014poverty}). More recent works have approached the stagnation of the GDP focusing on the structure of the countries' economic output rather than on their production volumes. For example, \cite{pugliese2017complex} showed that not only the current level of income of a country but the level of knowledge embodied in its economic activities, also known as economic complexity (see \citealp{hausmann2014atlas}), determines the minimum requirements for it to escape or to do not fall into a neoclassical poverty trap. More precisely, a country will not stagnate in an income level as long as a specific combination of its GDP and   economic complexity surpasses a given threshold.\\

Nonetheless, the neoclassical understanding of poverty traps as the stagnation of GDP growth seems insufficient to address the multidimensional features inherent to its causes. Despite economic growth is good for the poor, it is not enough as policy making \citep{vskare2016poverty}. The identification of low GDP with poverty is problematic since it assumes an even income distribution within a country's population. Thus, the shortcomings of using the GDP as a measure of poverty are (at least) its incapability to establish what fraction of the population lives in extreme poverty and to determine the evolution of this fraction over  time. On the one hand, the studies of neoclassical poverty traps usually involve the usage of aggregative models that overlook the implications of the heterogeneous structure of the economic activities operating in a region \citep{hausmann2011network}, i.e., activities with different levels of sophistication interacting in a spatial unit that shapes the evolution of the local economies. Hence, although the studies mentioned above propose valid schemes to fight poverty, such as improving institutions quality, investing in human or physical capital through education or R\&D, these solutions are themselves aggregative and do not provide either clear paths for the economies to escape from poverty traps or specific directions for economic policy development to prevent them. \\

The present study aims to provide a new quantitative framework to identify the risk that an economic region may face falling into a poverty trap, based on its current productive structure, i.e., we assess how the current set of economic activities of a spatial unit can explain the prevalence of poverty levels in it.  We draw the link between poverty and productive regional structure combining broadly used poverty measures (e.g., the headcount index) and the theory of economic complexity introduced by Hausmann, Hidalgo, and other collaborators \citep{hidalgo2007product}. Our analysis relies on the so-called product space (PS), i.e., a network that relates products through the shared knowledge or capabilities, also known as know-how, required to be produced \citep{hidalgo2007product}. Since the PS can be used to understand the productive structure of a spatial unit, it serves as an instrument to predict quiescence traps, which implicitly are related to poverty. A quiescence trap is the tendency of a region to remain in the production of only a set of goods, obstructing its ability to innovate \citep{hidalgo2007product}. We link poverty to production, constructing an index that reflects the level of association between each economic activity and the poverty in the economies that produce it, following the ideas of \cite{hartmann2017linking}. With this index at hand, we relate the quiescence traps with the poverty traps showing that specializing in a specific production basket not only prevents a region from entering into new activities but dooms it to remain in the production of goods that reinforce the cycle of poverty.\\

\section{Literature review}

This section is devoted to reviewing briefly three approaches that address poverty traps: (1) from neoclassical or aggreative viewpoint, (2) from an Economic Complexity perspective, and (3) from relatedness theory. First, we start pointing out the main aspects and shortcomings of the neoclassical definitions of poverty traps. Then, with (2) and (3), we present the main ideas that underpins our work. With this ideas at hand, we will present a novel framework to detect and measure poverty traps in the light of relatedness theory.\\

Regarding economic growth and development, poverty traps have been understood in terms of the current state of the countries in contrast with the performance of the global economy, i.e., it is approached from a comparative perspective. In this realm, explanations have been provided in an aggregative way, and the question in which the theory orbits is why some countries remain reinforcing their cycle of poverty. Several streams of literature have intended to answer this question. Neoclassical theory of growth (\citealp{solow1956contribution}; \citealp{romer1986increasing}; \citealp{lucas1988mechanics}) explains growth through capital accumulation, which determines the long run equilibrium that a country can achieve. From a mathematical point of view, growth and development have also been microfounded and conceptualized as multiple equilibria models. Here, poverty traps are stable equilibria that yield capital levels per capita below a given threshold. The type of equilibria attained by a country classifies them in two categories: ones stagnating its GDP below a poverty threshold and others growing  sustainably ( \citealp{romer1986increasing}; \citealp{lucas1988mechanics}; \citealp{azariadis1990threshold}; \citealp{durlauf1996theory}). In summary, poverty traps are associated with a low income and capital accumulation equilibrium (\citealp{barro2004economic}; \citealp{azariadis2005poverty}; \citealp{kraay2014poverty}). Despite aggregative models were the available tools to explain poverty, the explanation of poverty traps as a systematic failure in the accumulation of human and physical capital is rather indirect and does not capture the reality and multidimensionality of the phenomenon  \citep{hausmann2011network}.\\

Recent studies have borrowed tools from complexity and relatedness theory to explain poverty traps. These theories advocate for understanding the evolution of the economies through the analysis of the structure of their economic output (see \citealp{hidalgo2007product}; \citealp{hausmann2011network}). Complexity theory, for instance, defines a measure of sophistication for products in terms of the amount of know-how embodied in their production, i.e., the product complexity index (PCI), and a related measure for countries in terms of the availability of know-how within them, i.e., the economic complexity index (ECI) \citep{hausmann2014atlas}. The ECI, for example, has been proved to be a strong predictor of economic growth \citep{hausmann2014atlas}, and the PCI has been set as a new benchmark to define target industries to develop in countries and regions to pursue their economic growth \citep{balland2019smart}. \cite{pugliese2017complex} improved the neoclassical understanding of poverty traps introducing complexity as a new dimension. The addition of complexity was motivated by the substantial reduction of poverty of some Asian countries (also defined as Asian Tigers) that in the '60s had lower levels of GDP than some South American countries, which nowadays remain in poverty. These findings suggested that the neoclassical theory for developing countries did not follow the empirical rules since the use of GDP as an only state variable homogenizes countries as same objects that follow certain growth paths. Therefore, considering complexity allows to capture and predict the role of industrialization patterns in growth through capabilities of countries, which combined with certain level of income, can foster an economy or let it remain in poverty. Following this line of thought, \cite{pugliese2017complex} measure the possibility of a country to escape of (neoclassical) poverty traps through a convex combination of the logarithms of GDP per capita and complexity, i.e., a country needs a minimum level of complexity combined with a minimum level of GDP to guarantee the escape from a poverty trap.\\

On the other hand, some works in relatedness theory have also linked production with neoclassical poverty traps directly and indirectly. Directly, \cite{brummitt2017contagious} developed a based agent model using economic complexity and an input-output network to explain how big-pushes in technologies impact poverty, i.e., large jumps in the production of complex goods arise disruptions or coordination problems carrying an economy into a poverty trap in terms of functional agents and inputs. \cite{hausmann2011network} indirectly address poverty traps through the notion of quiescence traps, which is when some economies are unable to diversify significantly their exports basket because of their current availability of capabilities or knowledge. This amounts to saying that countries' initial endowment of capabilities could constraint drastically their innovation and production possibilities. When countries fall into quiescence traps, they remain stuck in the production of goods without innovation possibilities due to cognitive distances to produce other goods. \\

There are theories related to quiescence traps which have provided evidence that links productive structures with poverty and poverty traps. A well documented example of this are countries that have a \textit{curse} due to the intensive production of goods associated to natural resources  \citep{sachs2001curse}, i.e., countries that have remained in high levels of poverty (in terms of the headcount measure) because of their abundance of resources. For instance, recent studies \citep{apergis2018poverty} have shown a strong association between the production of petroleum, a peripheral product in the PS \citep{hidalgo2007product}, and high levels of the poverty (measured through the headcount index). Our work is motivated by the natural question that arises in the context of the quiescence traps: the inability of a country to diversify (i.e., to remain trapped in certain zones of the PS) has a direct implication in its opportunities to decrease its poverty levels?\\

With this work, we aim to deliver a new approach to understand poverty traps from the relatedness theory, using a poverty measure directly. Despite works as  \cite{brummitt2017contagious} use relatedness to explain poverty traps through coordination problems, our approach seeks to provide from a productive perspective a diagnose of countries' status in poverty and the nuances of it. Thus, since poverty traps have been addressed at an aggregate level or have been studied partially as the result of some economic phenomena, e.g., coordination problems, we aim to contribute by proposing a quantitative method to detect, even an early or utter detection, poverty traps. For this purpose, we will start assigning poverty to goods following the ideas of \cite{hartmann2017linking}. After linking poverty and production, we can asses the current position of a country in the PS in terms of the overall poverty of the goods in which it is located at. With this measure at hand, we study in what extent countries that remain producing the same basket of products (quiescence traps) that are also associated with high poverty levels (in average) have remained poor for long periods (poverty traps). 

 
\section{Poverty at product scale}

In this section, we present our strategy to measure poverty traps. Our starting point consists in linking production and poverty by mapping the headcount index explained in \cite{makoka2005poverty} to the PS introduced in \cite{hidalgo2007product}. This approach is motivated by the findings in \cite{hausmann2014atlas}, \cite{hausmann2007you}, and \cite{hartmann2017linking}, where quantities of economic interest such as income levels (GDP) or inequality (GINI index) are mapped from countries to products to explore in what extent the economic growth or the evolution of the inequality of countries can be explained by the structure of their exports basket. Along similar lines of reasoning, we derive two measures that assign poverty levels (or coefficients) to tradable goods based on the headcount index. The ultimate goal of this procedure is to show that countries specialized in the production of goods with high poverty levels are prone to linger in poverty for long periods.\\

This section is structured as follows: first, we start reviewing the constituents of our model, i.e., the PS and the headcount index. Second, we define the product poverty index (PPI) as the projection of the headcount index from countries to products. Third, we use the PPI and the PS network as inputs to construct the Eigenpoverty index that captures the long-run poverty levels of products, given their position in the network. Fourth, we introduce two measures to capture the potential of a country to improve its current poverty values based on the PPI and the Eigenpoverty index. Lastly, we define stagnation of countries using the headcount index.

\subsection{The product space}

The PS was devised in \cite{hidalgo2007product} to measure the relatedness between tradable goods in the world market, in terms of their probability of being co-exported with comparative advantage. In this case, the specialization is measured through the revealed comparative advantage (RCA) index  \citep{balassa1964purchasing} and is given by

\begin{equation}
   \text{ \textbf{RCA}}_{c,p} = \frac{x_{cp}/\Sigma_p x_{cp}}{\Sigma_c x_{cp}/\Sigma_{c,p}x_{cp}},
\end{equation}

where $x_{cp}$ is the amount of the good $p$ exported by the country $c$, and the sums run over all possible countries and products. Given these indices for each good and country, \cite{hidalgo2007product} defines the intensity of relatedness between goods $l$ and $m$ as:

\begin{equation}\label{eq links PS}
    y_{lm} = min\{\mathbb{P}( \text{ \textbf{RCA}}_{m,p} >1|  \text{ \textbf{RCA}}_{l,p} > 1),\mathbb{P}( \text{ \textbf{RCA}}_{l,p} >1|  \text{ \textbf{RCA}}_{m,p} > 1)\}.
\end{equation}
 
That is, the minimum of the conditional probability of obtaining RCA in a good, given that the country already has RCA in the other good (the minimum is taken so that the weight is undirected). Hence, the PS is defined as the undirected network of available goods in the
world economy with links weighted by the numbers $y_{lm}$. Since the PS is a graph without loops, we set $y_{ll}=0$. Given their relevance for some computations in this work, we also consider the normalized weights (or density ratios)

\begin{equation}\label{eq normalized weight}
    \phi_{pq}= \frac{y_{pq}}{\sum_{r} y_{pr}}.
\end{equation}

\subsection{Headcount measure and product poverty index}

For our analysis, we have used the headcount index as a measure of poverty. This measure considers the portion of the population that lives in extreme poverty, i.e.,  with less than  1.90 \$USD (2011 PPP) per day, which is the international poverty line established by the Wold Bank. The headcount index $H_c$ for the country $c$ is calculated as 

$$ H_c =  \frac{1}{N_c} \sum_{i=1}^{N_c}I(y_{i,c}<z) = \frac{N_{pov,c}}{N_c}$$ 

where $N_c$ is the total population of the country, $z$ is the poverty line, $y_{i,c}$ is the welfare indicator, i.e., consumption or income per capita in the country, $I(.)$ takes the value of $1$ when income, consumption falls below the poverty line and $0$ otherwise; and $N_{pov,c}$ is the number of poor people in the country. \\

We use the headcount index in the same spirit of the export sophistication measures such as the Product Complexity Index \cite{hausmann2014atlas}, Product GINI Index \cite{hartmann2017linking}, and the PRODY \cite{hausmann2007you}. In our case, we argue that decomposing the poverty at product level allows improving the understanding of the link between the current productive structure of a country and the possible causes of the perpetuation of poverty. A clear link between economic stagnation and the structure of economic output is the  quiescence traps, already discussed in the introduction.
The PS by itself explains (partially), in terms of its topological and probabilistic structure, why some countries struggle to diversify into new products and remain producing the same basket of goods for long periods. However, in terms of poverty, there are several nuances that are not captured solely by the PS, since low economic growth does not amount to be in a poverty trap as we discussed in the introduction (see \citealp{vskare2016poverty}). Thus, linking poverty to the PS introduces an extra layer of classification that contributes to determining to what extent poverty traps coincide with quiescence traps and assess how dangerous a quiescence trap could be.\\

Then, to relate poverty to the PS, we introduce the PPI as 

\begin{equation}\label{eq PPI}
    PPI_p = \frac{1}{Q_p}\sum_c M_{cp}s_{cp}H_c,
\end{equation}

where $s_{cp}= \frac{x_{cp}}{\sum_{p} x_{cp}}$ is the share of the product $p$ in the basket of exports of country $c$, $x_{cp}$ is the amount of exports of $c$ in the good $p$, $M_{cp}$ is a binary variable that takes the value one whenever $RCA_{cp}>1$ and zero otherwise, and $Q_p=\sum_c M_{cp}s_{cp}$.\\

If we regard the products as vectors or combinations capabilities as in \cite{hidalgo2007product} or \cite{hidalgo2009building}, we can think of the PPI as a measure of the amount of poverty associated with the capabilities integrating each product. These capabilities, as put forward in \cite{hidalgo2007product}, are regarded as the requirements in the broader sense for the production of a good, e.g., infrastructure, institutional conditions, know-how, tacit knowledge, and all the underlying scaffolding that supports such production. Several of these capabilities, as we discussed in the introduction, are associated with the various and intertwined causes of poverty. As an instance of this, it has been extensively studied (see, e.g., \citealp{acemoglu2001colonial}; \citealp{acemoglu2005institutions}; \citealp{tadei2018long}) that post-colonial economies where extractive institutions were strongly rooted ended up specializing in the production of mineral and agricultural goods such as gold, coffee, cocoa or cotton (see also Table \ref{table:10pooresteign}). Another example is the so-called curse of the natural resources, i.e., countries with great abundance of natural resources (like minerals or oil) that tend to grow more slowly than resource-poor countries \citep{sachs2001curse}. In this latter example, it has been shown that the continuous extraction and export of certain goods (such as fossil energy resources) appear to increase the poverty (in terms of the headcount measure) of countries with relative abundance of these goods \citep{apergis2018poverty}. \\

\begin{figure}[h!]
    \centering
    \begin{subfigure}[h]{0.49\textwidth}
    \centering
        \includegraphics[scale=0.35]{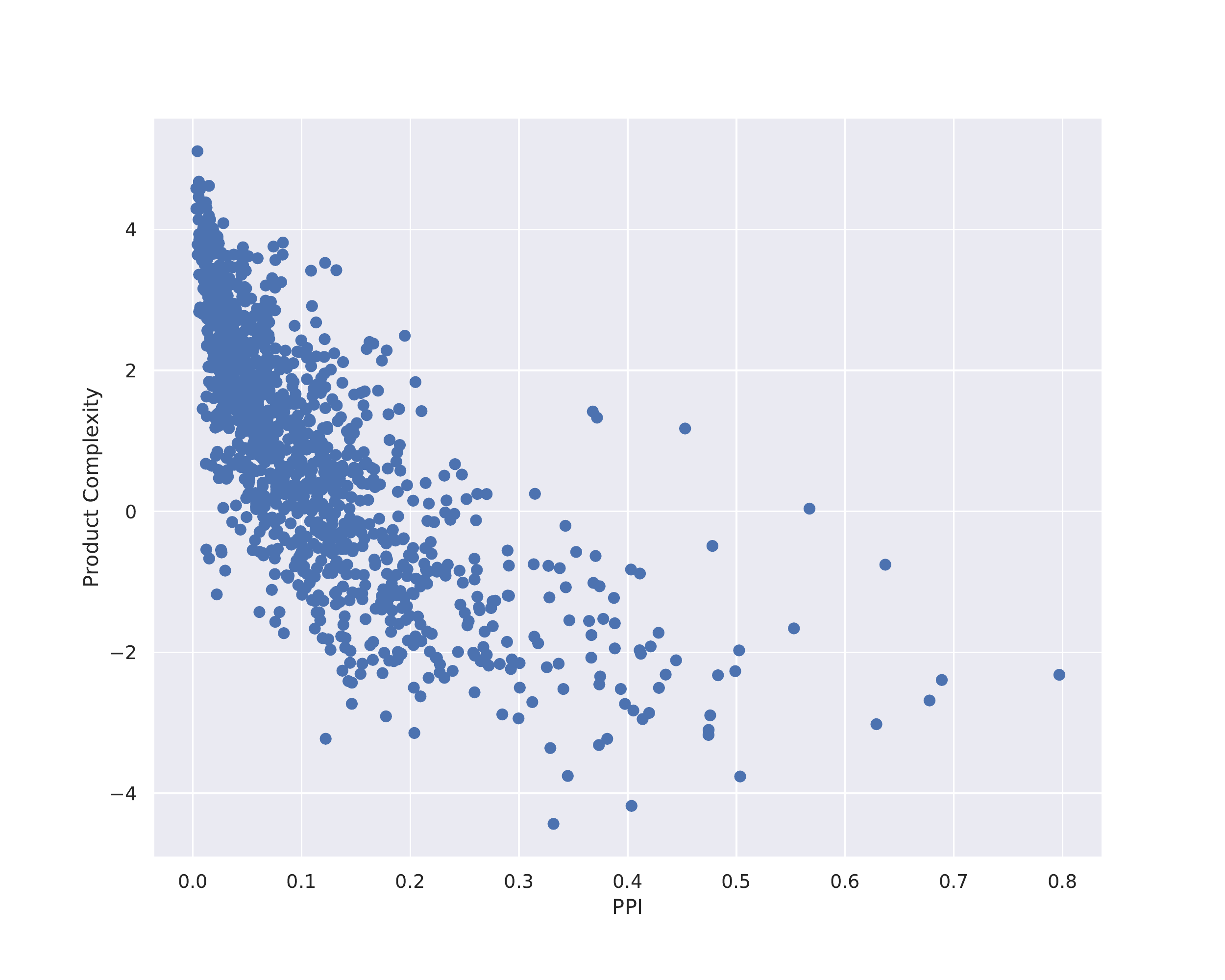}
        \caption{PPI vs. Product Complexity in 1995-2010}
        \label{ppi_vs_pc}
    \end{subfigure}
    \hfill
    \begin{subfigure}[h]{0.49\textwidth}
    \centering
        \includegraphics[scale=0.35]{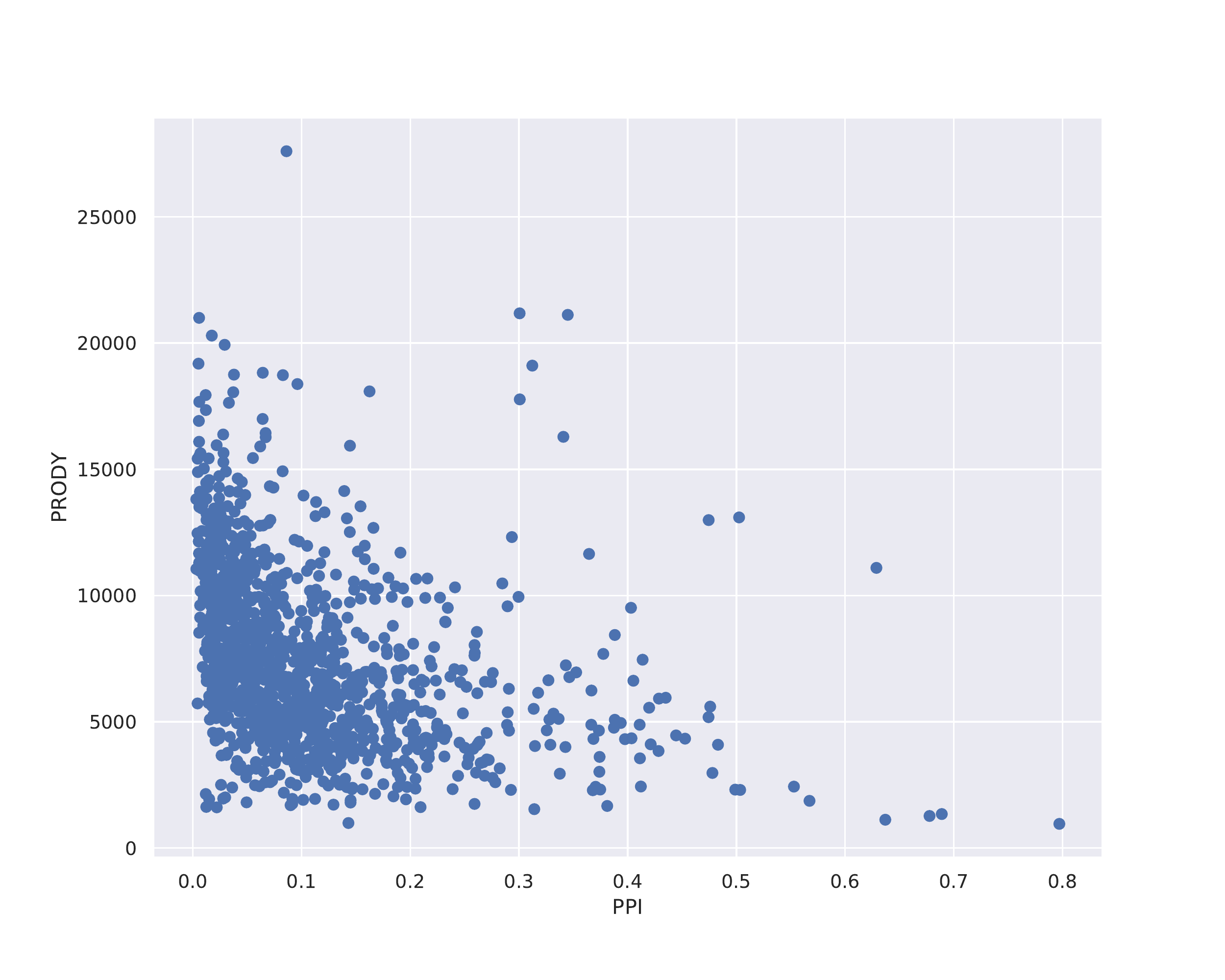}
        \caption{PPI vs. PRODY in 1995-2010}
        \label{ppi_vs_prody}
    \end{subfigure}
    \caption{Comparison of PPI with other product structured measures.}
\end{figure}

The arguments and examples that have served us to justify the association between poverty and production are closely related with evidence that has supported the introduction of other export sophistication measures  like the Product Complexity Index \citep{hausmann2014atlas} or the PRODY \citep{hausmann2007you}. For this reason, we compare the PPI with the two aforementioned measures to understand the differences and similarities between them and our measure of poverty. Figure \ref{ppi_vs_pc} depicts the relationship between Product Complexity and PPI per product, showing a negative correlation. This relationship indicates that the sophistication of the capabilities (know-how) is negatively correlated with the poverty carried by them, which goes in line with our standpoint and will be further discussed in Section \ref{sec:robustness}. Moreover, Figure \ref{ppi_vs_pc} suggests a quadratic relationship between PPI and PCI.  On the other hand, Figure \ref{ppi_vs_prody} relates PPI with the PRODY of \cite{hausmann2007you}. The PRODY captures income at product level \citep{hausmann2007you}. However, the liaison between the PRODY and the PPI is less clear than in the case of the Product Complexity Index and the PPI. This ambiguity shows that, even at product level, a certain degree of poverty can coexist with different levels of income.\\

\begin{figure}[h!]
    \centering
    \begin{subfigure}[h]{0.49\textwidth}
    \centering
        \includegraphics[scale=0.4]{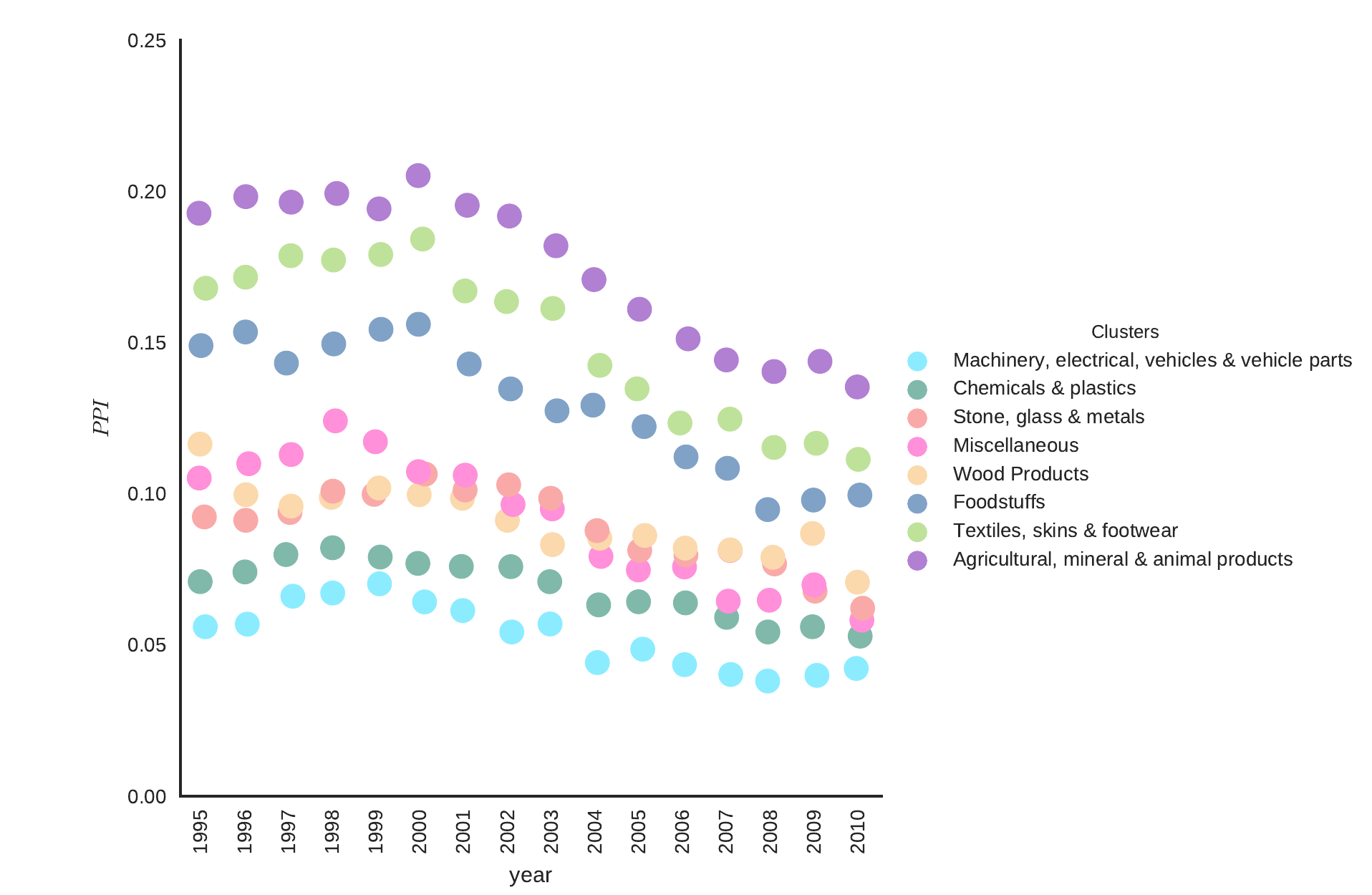}
        \caption{Ranking of industrial clusters 1995-2010}
        \label{ppi_rank_clust}
    \end{subfigure}
    \hfill
    \begin{subfigure}[h]{0.49\textwidth}
    \centering
        \includegraphics[scale=0.3]{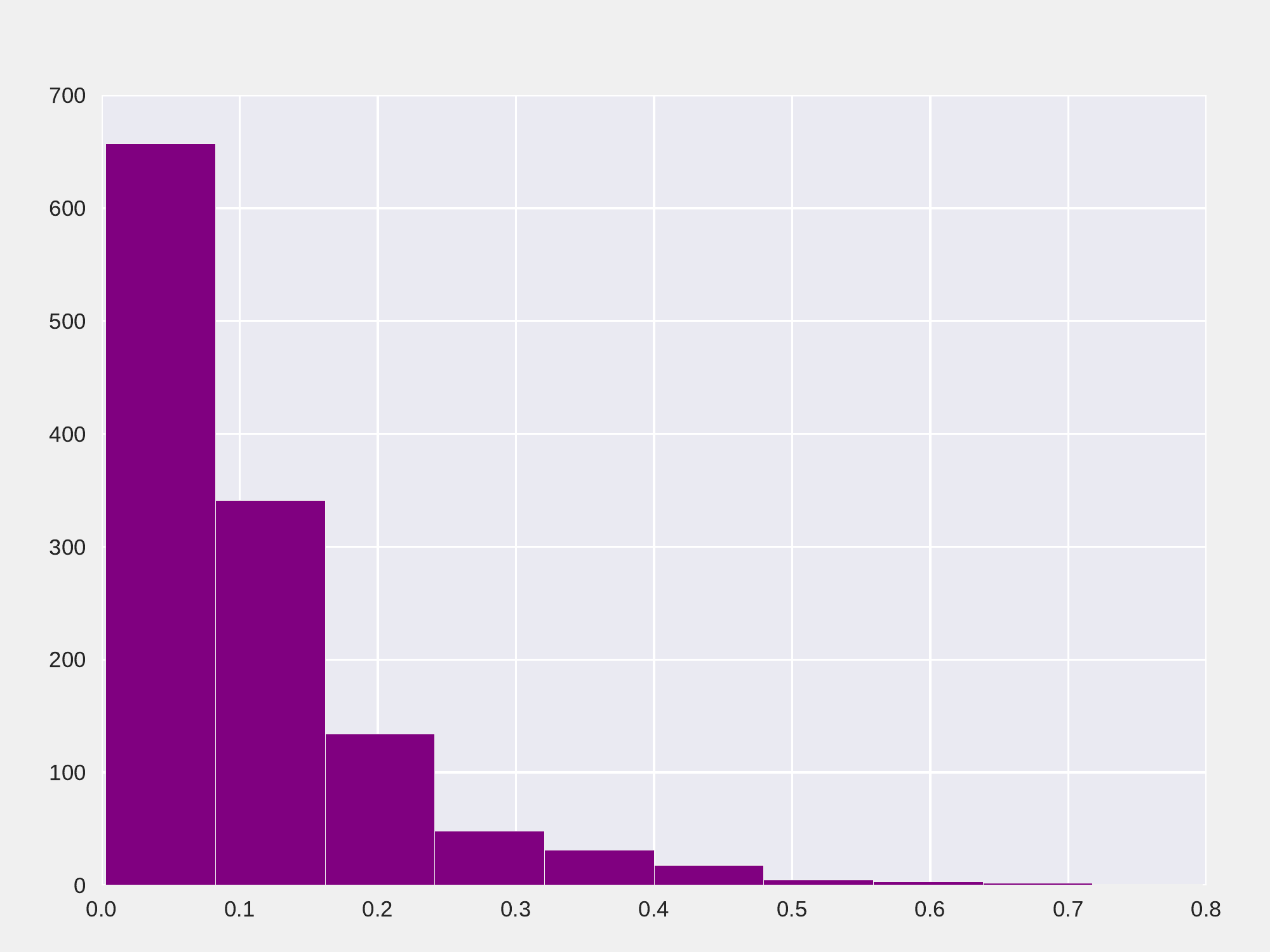}
        \caption{Distribution of PPI 1995-2010}
        \label{ppi_dist}
    \end{subfigure}
    \caption{PPI by industrial clusters its distribution}
\end{figure}

For the sake of clarity, we group the products in industrial clusters as depicted in Figure \ref{ppi_rank_clust}. In this graph, we give a closer look at the stability of the PPI per industrial cluster. Agricultural, mineral \& animal products; textiles, skins \& footwear; and foodstuffs are the industries that throughout time are associated with higher poverty levels, while machinery, electrical, vehicles \& vehicle parts; and chemicals \& plastics are associated with lower levels of poverty. On the other hand, Figure \ref{ppi_dist} exhibits the distribution of the PPI, where we note that it is right-skewed, indicating the existence of a small set of goods that are significantly poorer than the rest. We present more detailed information about these latter goods in Table \ref{table:10poorestppi}, where we show the ten poorest products in 2010, ranked by PPI, together with their product complexity index, and the three more prevalent producers of these goods. Let us notice that those products are primarily of agricultural and mineral origin and that their product complexity (PCI) is negative.

\begin{table}[htbp]
      \centering
         \scalebox{0.46}{
        \begin{tabular}{|c|l|l|l|}

\hline
\textbf{Product}                                  & \multicolumn{1}{c|}{\textbf{PPI}} &  \multicolumn{1}{c|}{\textbf{PCI}} & \multicolumn{1}{c|}{\textbf{Country}} \\ \hline
\multirow{3}{*}{Cobalt Ore}                       & \multirow{3}{*}{0.83021191}                & \multirow{3}{*}{-3.1659999}       & Congo, DR                             \\
  &                                   &                                  & Zambia                                \\
 & & & Russia                                \\ \hline
\multirow{3}{*}{Vanilla}                          & \multirow{3}{*}{0.733408008}                & \multirow{3}{*}{-1.5049}          & Madagascar                            \\
 &                                                                               &                                   & France                                \\
 &                                                                               &                                   & Germany                               \\ \hline
\multirow{3}{*}{Cobalt}                           & \multirow{3}{*}{0.726402948}               & \multirow{3}{*}{-0.74194002}      & Congo, DR                             \\
 &                                                                              &                                   & Zambia                                \\
  &                                                                              &                                   & Finland                               \\ \hline
\multirow{3}{*}{Agave}                            & \multirow{3}{*}{0.668886363}               & \multirow{3}{*}{-2.8546}          & Brazil                                \\
                                                                                    &                                            &                                   & Kenya                                 \\
                                                                                     &                                            &                                   & Madagascar                            \\ \hline
\multirow{3}{*}{Cloves}                           & \multirow{3}{*}{0.662467203}               & \multirow{3}{*}{-1.8163}          & Madagascar                            \\
                                                                                     &                                            &                                   & Sri Lanka                             \\
                                                                                     &                                            &                                   & Indonesia                             \\ \hline
\multirow{3}{*}{Vegetable Plaiting Materials}     & \multirow{3}{*}{0.558597641}               & \multirow{3}{*}{-1.0895}          & China                                 \\
                                                                                     &                                            &                                   & Indonesia                             \\
                                                  &                                                                               &                                   & Argentina                             \\ \hline
\multirow{3}{*}{Graphite}                         & \multirow{3}{*}{0.521807796}               & \multirow{3}{*}{-0.1036}          & China                                 \\
                                                  &                                                                               &                                   & Brazil                                \\
                                                  &                                                                               &                                   & Japan                                 \\ \hline
\multirow{3}{*}{Prepared Cotton}                  & \multirow{3}{*}{0.46906604}                      & \multirow{3}{*}{-3.3611}          & Mali                                  \\
                                                  &                                                                               &                                   & Malaysia                              \\
                                                  &                                                                               &                                   & Pakistan                              \\ \hline
\multirow{3}{*}{Aluminium Ore}                    & \multirow{3}{*}{0.447409404}               & \multirow{3}{*}{-2.8225}          & Indonesia                             \\
                                                                                     &                                            &                                   & Guinea                                \\
                                                  &                                                                               &                                   & Brazil                                \\ \hline
\multirow{3}{*}{Coconuts. Brazil Nuts \& Cashews} & \multirow{3}{*}{0.309893018}               & \multirow{3}{*}{-4.1026001}       & Vietnam                               \\
                                                  &                                                                               &                                   & India                                 \\
                                                  &                                                                               &                                   & Cote dIvoire                          \\ \hline
\end{tabular}
\caption{10 poorest products ranked by PPI in 2010.}
\label{table:10poorestppi}}
\end{table}

\begin{figure}[h!]
    \centering
    \includegraphics[scale=0.6]{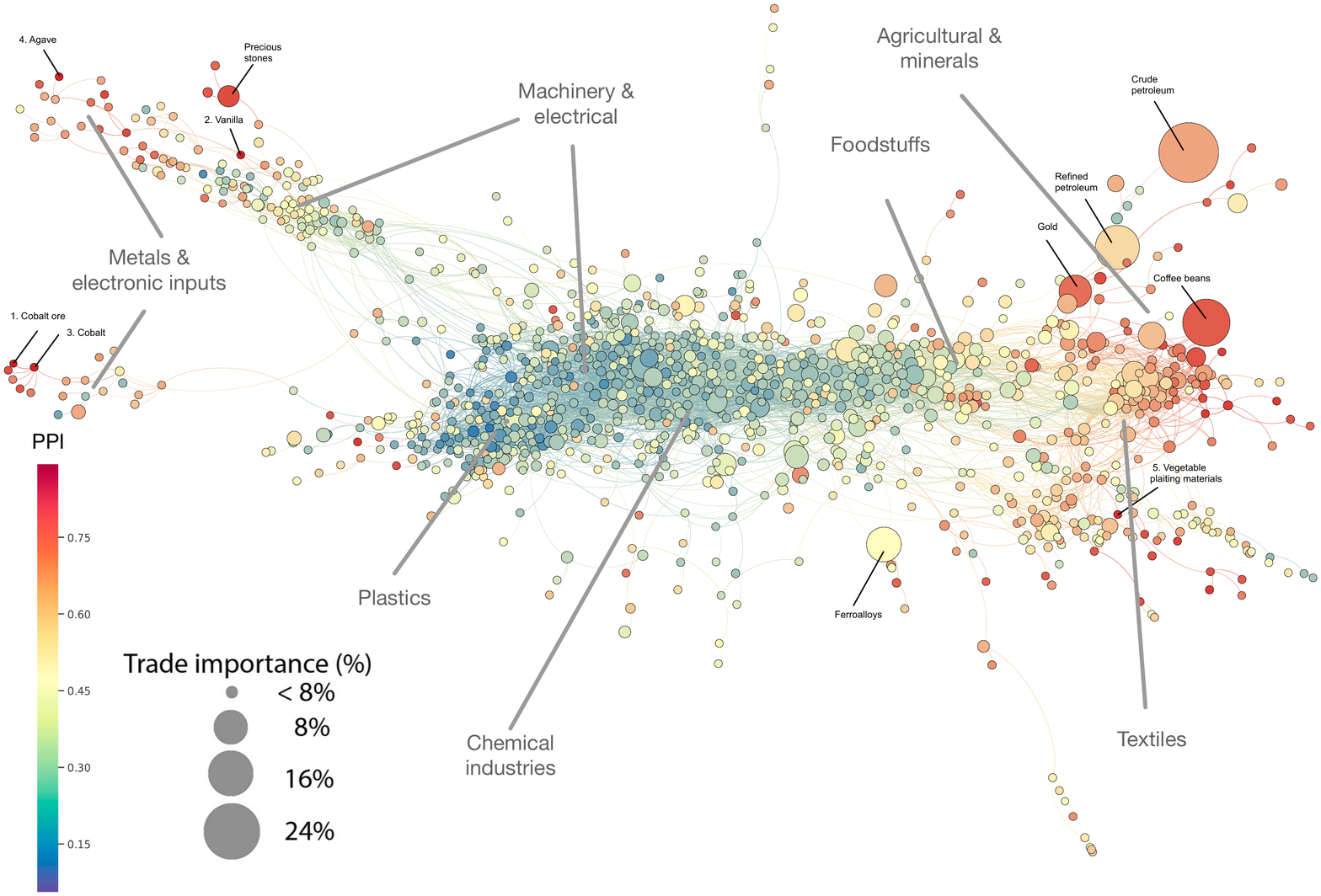}
    \caption{PPI distribution on the PS}
    \label{ppi_graph}
\end{figure}

Figure \ref{ppi_graph} depicts the distribution of the PPI on the PS. For this construction, we use the average the PPI in 1995-2010. To improve the visualization, we rescale the distribution of the PPI taking the square root of its values. Additionally, the size of the nodes represents the average trade importance of each good. Additionally, we have remarked the five poorest goods and those that are conspicuous for their trade importance. It is noticeable that the periphery of the PS is mostly populated by goods associated with high poverty levels. These products are primarily in the industrial clusters of agriculture \& minerals, and textiles. There are also goods that belong to clusters like machinery \& electrical, plastics, and chemical industries, but that are of extractive nature and are associated to those clusters because of their role as input in many activities that also belongs to such clusters (e.g., certain stones or metals required extensively in the electrical industries end up belonging to this cluster). On the other hand, goods in the center of the PS tend to have lower levels of poverty, where the industries are of non-extractive nature. This heat-map serves as a basic yet informative descriptive that helps to locate zones in the PS of concentrated poverty, which can also help to diagnose whether an economy is into a poverty trap (see Section \ref{sec:caseofstudy}).

\subsection{Eigenpoverty index} \label{subsection:eigenpoverty}

 As it is defined, the PPI associates to each tradable product the current poverty levels of its principal producers. Given the dynamic component of our variable of interest (the stagnation of poverty), we introduce the Eigenpoverty index as a complementary measure to capture the potential poverty contribution of each product in the long run. We introduce the eigenpoverty as a modification of the well-known eigenvector centrality or Katz centrality \citep{jackson2010social}. We start defining the poverty reduction potential (PRP) of a product $p$ as $PRP_p=1-PPI_p$. The principle upon which the Katz prestige is build asserts that the prestige of a node in a network should be proportional to the sum of the prestige of its neighbors multiplied by the weight of the corresponding link. Keeping the the analogy with Katz prestige, we define the Eigenpoverty$'$ of a good $p$ ($E'_p$) as
 
  \begin{equation}\label{eq eigenpoverty}
 E'_p = \lambda PRP_p\sum_{q} \phi_{pq} E'_q,   
 \end{equation}
 
 where $\phi_{pq}$ is the normalized weight of the links of the PS introduced in equation \eqref{eq normalized weight}, and $\lambda>0$ is a constant. In contrast with Katz prestige we add the factor $PRP_p$, postulating that the ability of the product $p$ to gain prestige from its neighbors is conditioned by its potential to reduce poverty.
Since the Eigenpoverty index of all products is determined simultaneously, it is convenient to understand equation \eqref{eq eigenpoverty} in matrix form. Let $E'$ be the Eigenpoverty$'$ vector and $\Phi^*$ the normalized adjacency matrix of the PS corrected by the $PRP$, i.e.,

\begin{equation}\label{eq adjusted ad matrix}
    \Phi^* = \begin{pmatrix}
    PRP_{1} \phi_{11} &  PRP_{1} \phi_{12} & \cdots & PRP_{1} \phi_{1n}\\
     PRP_{2} \phi_{21} &  PRP_{2} \phi_{22} & \cdots & PRP_{2} \phi_{2n}\\
       \vdots &  \vdots  & \cdots & \vdots\\
           PRP_{n} \phi_{n1} &  PRP_{n} \phi_{n2} & \cdots & PRP_{n} \phi_{nn}\\
    \end{pmatrix}
\end{equation}

Hence, equation \eqref{eq eigenpoverty} reads in matrix forms as follows

\begin{equation}\label{eq eigenvalue problem}
    E'=\lambda \Phi^* E'.
\end{equation}

So, as in the case of Katz prestige, determining the value of $E$ amounts to solve an eigenvector-eigenvalue problem. Therefore, we define $E$ as the unique eigenvector associated with the largest eigenvalue of $\Phi^*$ such that $E_p \geq 0$ and $\sum_{p} E_p =1$. The existence and uniqueness of such eigenvector is guaranteed by Perron-Frobenius' theorem (see \citealp{jackson2010social}). However, since we used the $PRP$ as input to keep logical the analogy with Katz prestige, we define the Eigenpoverty of a good as 

\begin{equation}\label{eq eigenpovertyequation}
   E =  \textbf{1}- E'
\end{equation}

where \textbf{1} is a vector of ones.\\

From the perspective of the theory of Hausmann and Hidalgo, the conception of this index is motivated by the idea that the current poverty levels of each product (their PPI) are explained by the poverty inherent to the capabilities that conform it, and by the idea that the joint level of poverty associated with sets of products may increase or decrease depending on their relatedness structure (the PS). Thus, the long-run poverty level (or Eigenpoverty) of a good can be thought of as a correction of its current poverty level by the effect of the poverty levels of other products related to it (its neighbors in the PS). \\
\begin{figure}[h!]
    \begin{subfigure}[h]{0.49\textwidth}

        \centering
    \includegraphics[scale=0.35]{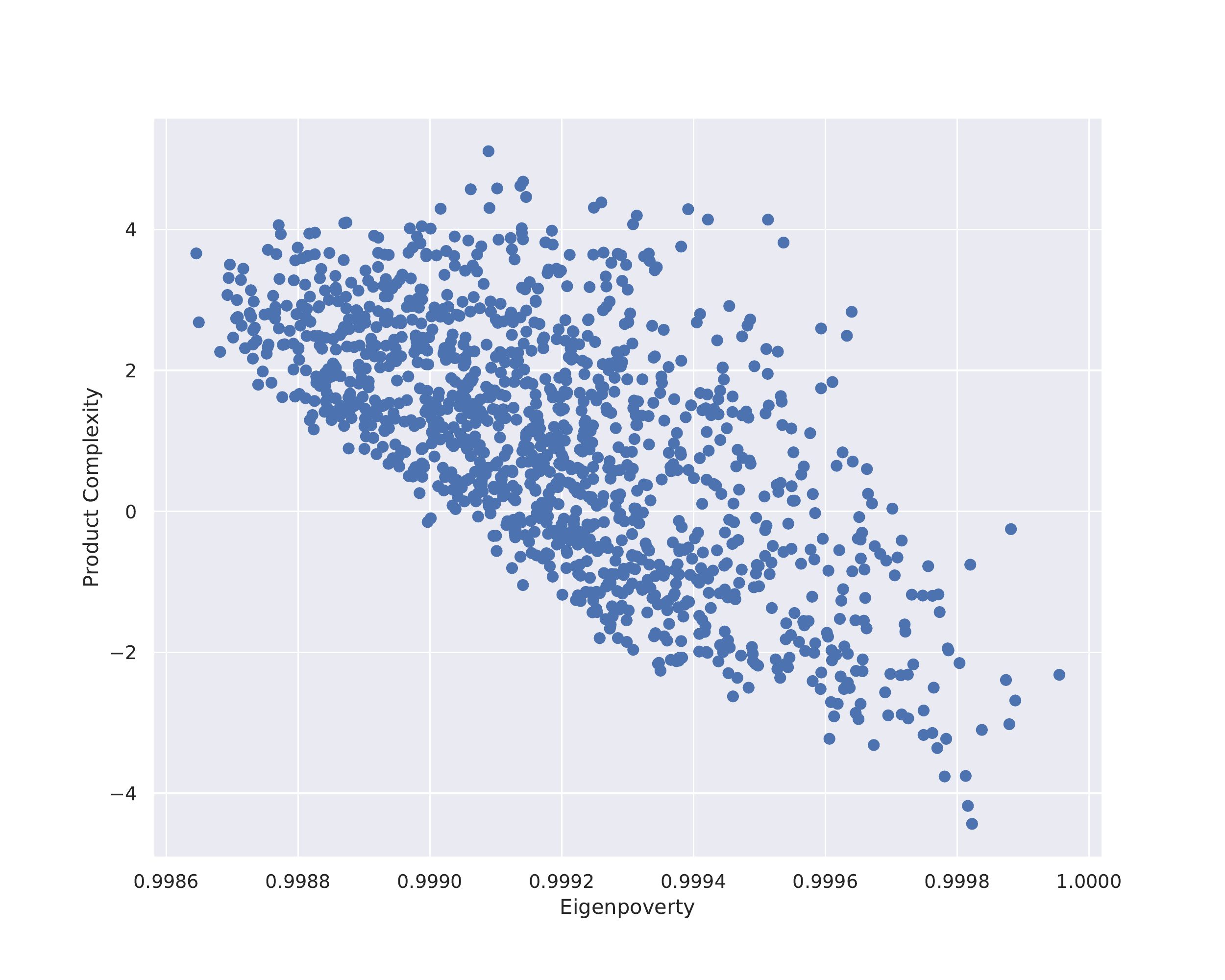}
    \caption{Eigenpoverty vs. Product Complexity for 1995-2010}
    \label{eigen_vs_compl}
    \end{subfigure}
    \hfill
    \begin{subfigure}[h]{0.49\textwidth}
    \centering
        \includegraphics[scale=0.35]{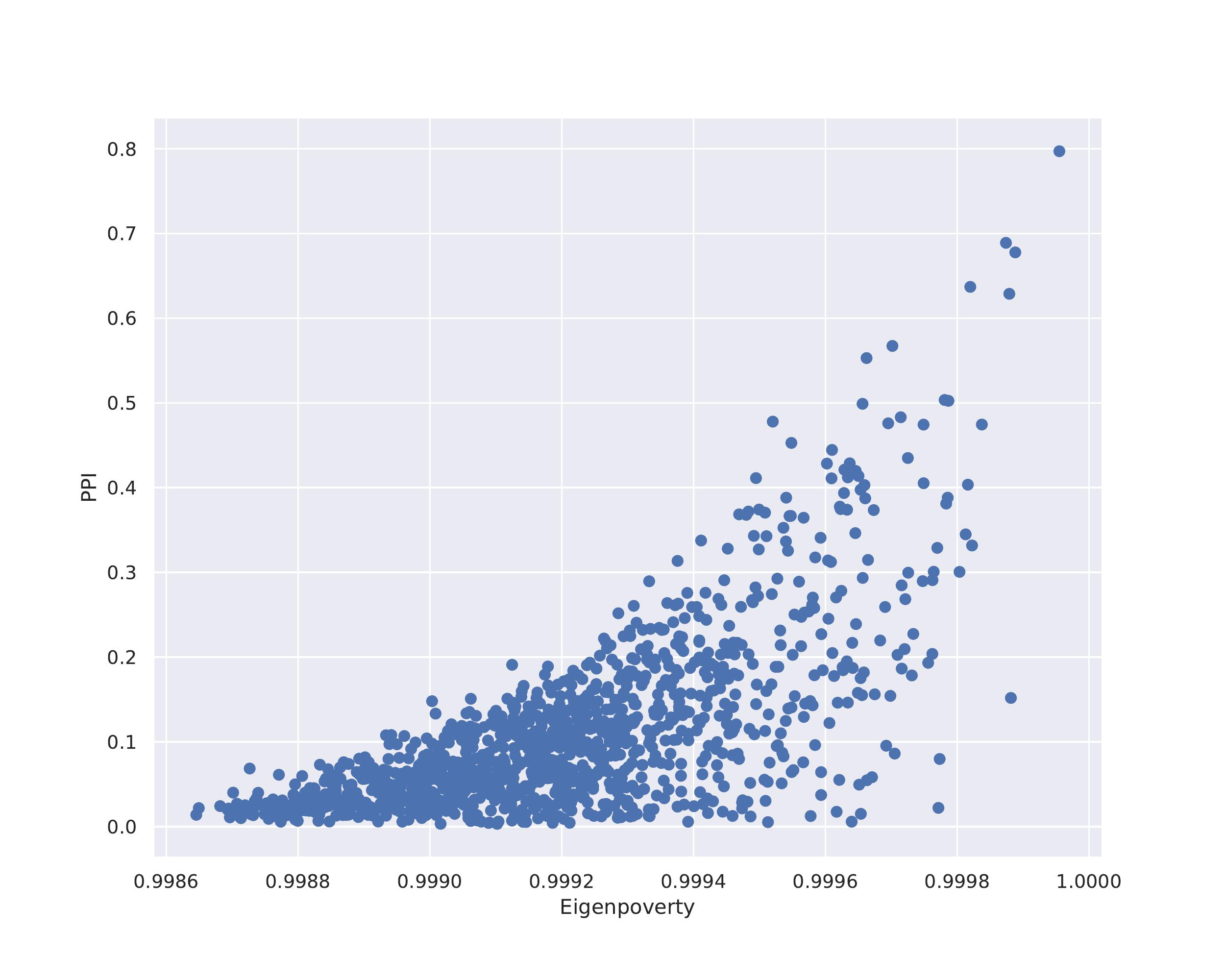}
        \caption{Eigenpoverty vs. PPI for 1995-2010}
        \label{eig_vs_ppi}
    \end{subfigure}
    \caption{Comparison of Eigenpoverty with other product structured measures.}
\end{figure}

Similarly, as in the previous section with the PPI, we study the behavior of the Eigenpoverty index comparing it with other two product measures, namely, the  PCI of \cite{hidalgo2009building} and the PPI. As in the case of the PPI, the Eigenpoverty index is negatively correlated with the PCI. Figure \ref{eigen_vs_compl} shows that this relationship seems linear, in contrast to the quadratic association depicted in Figure \ref{ppi_vs_pc}.  Figure \ref{eig_vs_ppi} compares both measures of poverty, the PPI, with the Eigenpoverty index.  Based on both visualizations, we point out that these two measures are not comparable since, for high values of Eigenpoverty, there is a high dispersion of the PCI. Thus, we might as well regard them as complementary indicators of poverty at product level. A possible explanation of this discrepancy is the global nature of the Eigenpoverty, which involves the topological information of the entire network. As an instance of this feature, the Eigenpoverty penalizes poorly connected zones of the PS (like the periphery), smoothing the transition from non-poor goods to poor from the core to the periphery, respectively.\\

\begin{figure}[h!]
    \centering
    \includegraphics[scale=0.6]{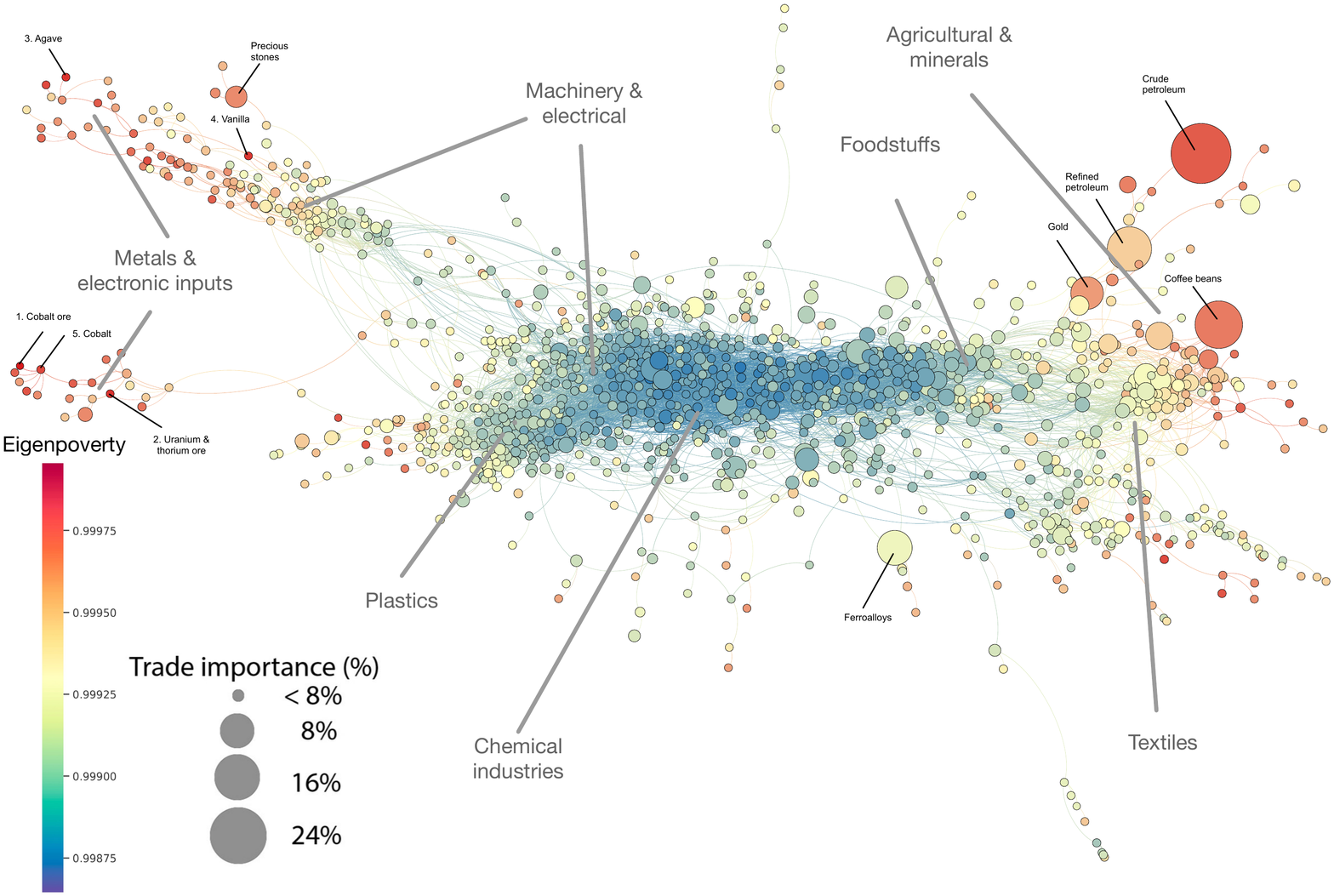}
    \caption{Eigenpoverty}
    \label{eigen_graph}
\end{figure}

Figure \ref{eigen_graph} depicts the distribution of the Eigenpoverty on the PS. Following the lines of the visualization of the PPI in Figure \ref{ppi_graph}, we label the five poorest products and the most important products according to their trade importance. We point out that the transition between non-poor products and poor products is more nuanced and clear in the case of the Eigenpoverty than in the case of the PPI, as mentioned above. For instance, in Figure \ref{ppi_graph}, the products coffee beans and gold are poorer than crude petroleum, but crude petroleum becomes poorer than gold and coffee beans in this new representation. This difference is mainly explained by the difference of connections of these goods, since crude petroleum is a leaf of the PS, whereas the other two products are (slightly) better connected. In Section \ref{sec:robustness}, we will revisit this discrepancy to argue, at the level of countries, that this obeys the fact that the PPI explains why some countries can remain in poverty through a period of time, i.e., short run poverty, whereas the Eigenpoverty predicts the stagnation of poverty in the long run. \\

   \begin{table}[!htb]
      \centering
        \scalebox{0.46}{
       \begin{tabular}{|c|l|l|l|l|}
\hline
\textbf{p}                                    & \textbf{Eigenpoverty}        & \multicolumn{1}{c|}{\textbf{PPI}} & \textbf{PCI}                 & \textbf{Country} \\ \hline
\multirow{3}{*}{Cobalt Ore}                   & \multirow{3}{*}{0.99996319}  & \multirow{3}{*}{0.83021191}       & \multirow{3}{*}{-3.1659999}  & Congo, DR        \\
                                              &                              &                                   &                              & Zambia           \\
                                              &                              &                                   &                              & Russia           \\ \hline
\multirow{3}{*}{Vanilla}                      & \multirow{3}{*}{0.99989424}  & \multirow{3}{*}{0.733408008}      & \multirow{3}{*}{-1.5049}     & Madagascar       \\
                                              &                              &                                   &                              & France           \\
                                              &                              &                                   &                              & Germany          \\ \hline
\multirow{3}{*}{Agave}                        & \multirow{3}{*}{0.999893645} & \multirow{3}{*}{0.668886363}      & \multirow{3}{*}{-2.8546}     & Brazil           \\
                                              &                              &                                   &                              & Kenya            \\
                                              &                              &                                   &                              & Madagascar       \\ \hline
\multirow{3}{*}{Cloves}                       & \multirow{3}{*}{0.999884319} & \multirow{3}{*}{0.662467203}      & \multirow{3}{*}{-1.8163}     & Madagascar       \\
                                              &                              &                                   &                              & Sri Lanka        \\
                                              &                              &                                   &                              & Indonesia        \\ \hline
\multirow{3}{*}{Cobalt}                       & \multirow{3}{*}{0.999867821} & \multirow{3}{*}{0.726402948}      & \multirow{3}{*}{-0.74194002} & Congo, DR        \\
                                              &                              &                                   &                              & Zambia           \\
                                              &                              &                                   &                              & Finland          \\ \hline
\multirow{3}{*}{Uranium \& Thorium Ore}       & \multirow{3}{*}{0.999867699} & \multirow{3}{*}{0.030407747}      & \multirow{3}{*}{-1.9313}     & Australia        \\
                                              &                              &                                   &                              & Kazakhstan       \\
                                              &                              &                                   &                              & South Africa     \\ \hline
\multirow{3}{*}{Cocoa Beans}                  & \multirow{3}{*}{0.999828119} & \multirow{3}{*}{0.346848453}      & \multirow{3}{*}{-4.1550999}  & Cote dIvoire     \\
                                              &                              &                                   &                              & Ghana            \\
                                              &                              &                                   &                              & Indonesia        \\ \hline
\multirow{3}{*}{Rubber}                       & \multirow{3}{*}{0.999814393} & \multirow{3}{*}{0.393659784}      & \multirow{3}{*}{-4.2079}     & Indonesia        \\
                                              &                              &                                   &                              & Thailand         \\
                                              &                              &                                   &                              & Malaysia         \\ \hline
\multirow{3}{*}{Jute \& Other Textile Fibers} & \multirow{3}{*}{0.999806797} & \multirow{3}{*}{0.316815732}      & \multirow{3}{*}{-4.7571998}  & Bangladesh       \\
                                              &                              &                                   &                              & India            \\
                                              &                              &                                   &                              & Kenya            \\ \hline
\multirow{3}{*}{Copra}                        & \multirow{3}{*}{0.999785461} & \multirow{3}{*}{0.297238499}      & \multirow{3}{*}{-3.4389}     & Egypt            \\
                                              &                              &                                   &                              & Papua New Guinea \\
                                              &                              &                                   &                              & Indonesia        \\ \hline

\end{tabular}}
\caption{10 poorest products ranked by eigenpoverty in 2010.}
\label{table:10pooresteign}
\end{table}

\begin{figure}[htbp]
    \centering
    \begin{subfigure}[h]{0.49\textwidth}
    \centering
        \includegraphics[scale=0.4]{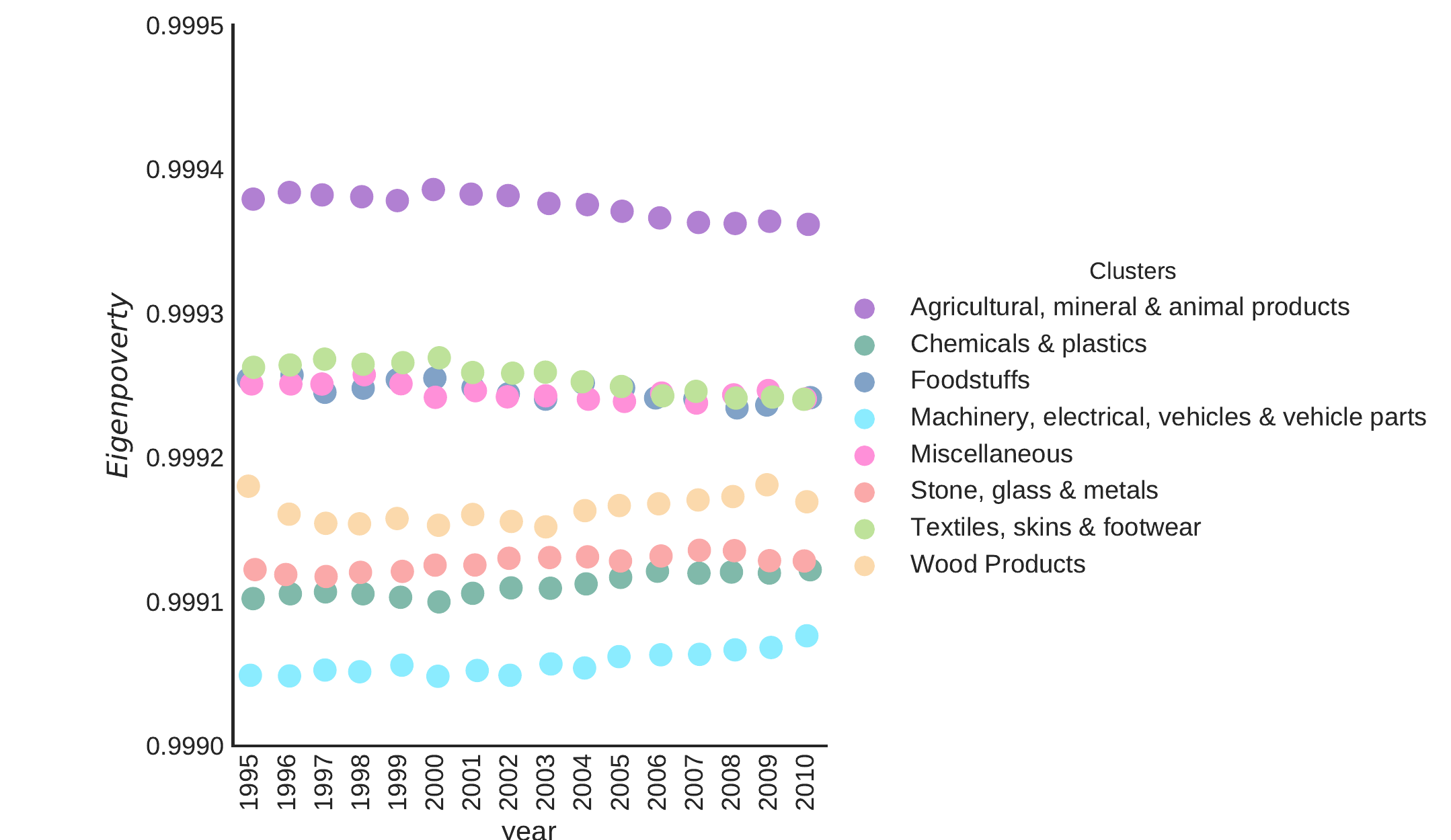}
        \caption{Ranking of industrial clusters 1995-2010}
        \label{eig_rank_clust}
    \end{subfigure}
    \hfill
    \begin{subfigure}[h]{0.49\textwidth}
    \centering
        \includegraphics[scale=0.3]{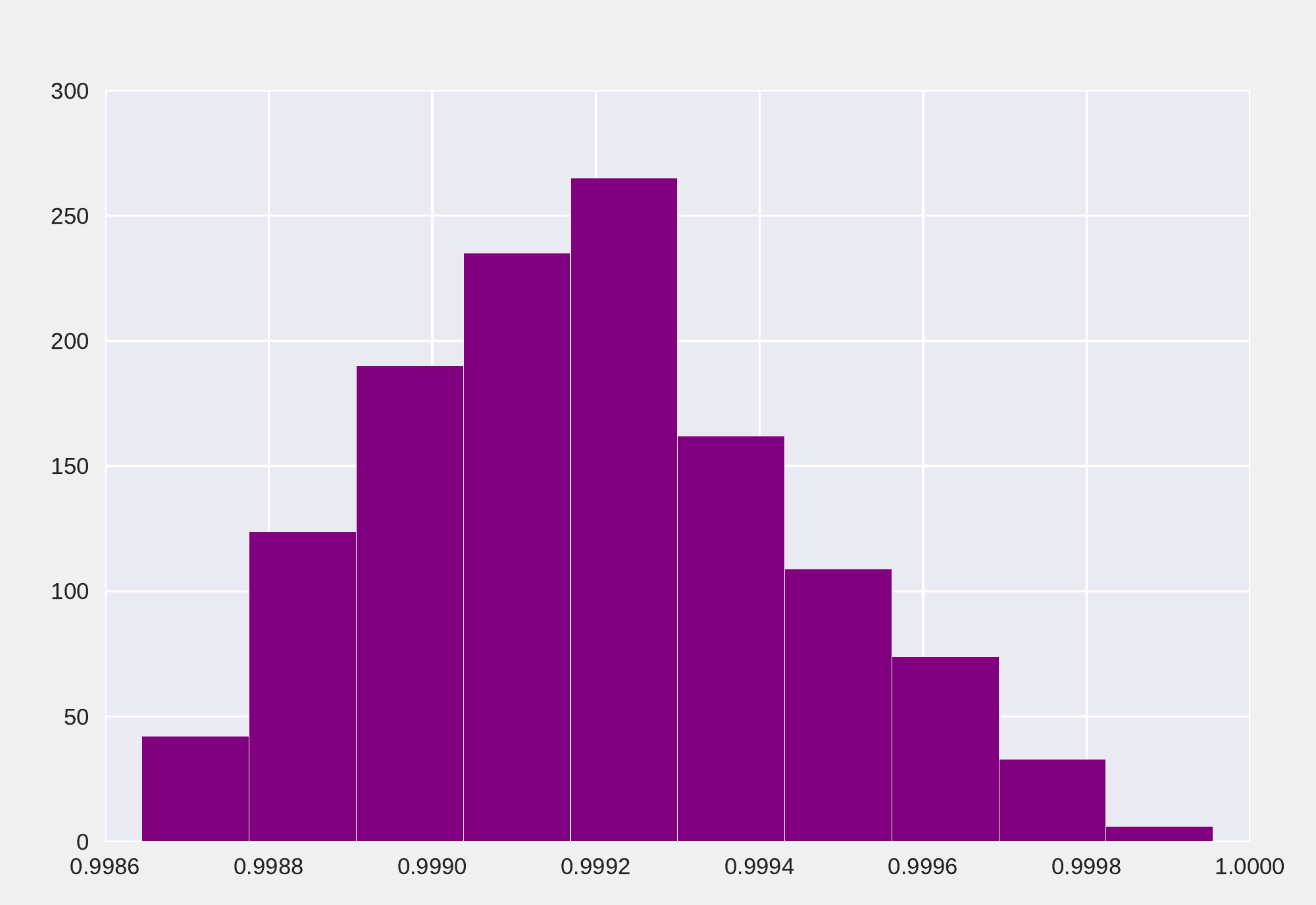}
        \caption{Distribution of Eigenpoverty 1995-2010}
        \label{eigen_dist}
    \end{subfigure}
    \caption{Eigenpoverty by industrial clusters and its distribution}
\end{figure}

In Figure \ref{eig_rank_clust}, we show the stability of the Eigenpoverty index by industrial clusters throughout time. As expected, the Eigenpoverty tends to be more stable than the PPI (see Figure \ref{ppi_rank_clust}) at cluster level. The industrial activities foodstuffs, textiles, skins \& footwear, and miscellaneous tend to have the same levels of poverty, and machinery\& electrical, vehicles \& vehicle parts and chemicals \& plastics have low levels of poverty. Additionally, agricultural \& mineral products are considerably poorer than the rest of the goods, relatively. On the other hand, Figure \ref{eigen_dist} depicts the distribution of the Eigenpoverty among goods. This distribution is more symmetric than the one depicted in Figure \ref{ppi_dist}, indicating that most goods, in the long run, have a middle level of poverty. This shape also corroborates the smooth distribution of the Eigenpoverty on the PS. \\

Table \ref{table:10pooresteign} shows the ten poorest products according to the Eigenpoverty index in 2010 with their corresponding PPI, PCI, and the three most important countries in exporting these products. The ranking of this two measures is similar, assigning high levels of poverty to mineral and agricultural goods such as cobalt ore or agave. However, there are some interesting differences, e.g., uranium \& thorium ore has a low level of PPI, but it is in the 6th poorest product according to the ranking of the Eigenpoverty index. These differences could indicate that, given the current distribution of market shares in the global economy, some goods may have an artificially deflated PPI, but, given their location in the PS, in the long-run their associated capabilities will tend to be poor.

\section{Data and methods}

\subsection{Data}
In this work, we consider the PS introduced in \citep{hidalgo2007product}. In its construction, we use trade data from the Centre d'\'Etudes Prospectives et d\'Informations Internationales (CEPII) for 1995-2010, including observations for 128 countries and 1240 products classified under the nomenclature of the Harmonized System at four digits (hs4). Analogously, we use the ECI and the PCI as presented in \cite{hausmann2014atlas}. We consider the whole network for our computations. Nonetheless, for visualization purposes in Figure \ref{ppi_graph} and Figure \ref{eigen_graph}, we filtered the network by the condition $y_{lm}>0.45$, with $y_{lm}$ introduced in equation \eqref{eq links PS}. Additionally, the resultant network was conveniently modified to introduce some nodes of central interest for our study.\\

The poverty data were obtained from the World Bank's portal PovcalNet, an online tool for global poverty monitoring, released in September 2020 (visit: {\color{blue}\url{http://iresearch.worldbank.org/PovcalNet/home.aspx})} with revised estimations of poverty based on Purchasing Power Parity (PPP) in constant 2011 US\$, including 166 countries, for which 28 are high-income economies. This information was available for the period 1981-2018. On the other hand, data at country level for GDP per capita at PPP in constant 2005 US\$ and population was taken from the World Bank's World Development Indicators; years of schooling came from Barro-Lee data set \citep{barro2013new}. The data that accounts for institutionality, such as corruption control, political stability and absence of violence/terrorism, government effectiveness, regulatory quality, and voice and accountability, were taken from the World Bank's Worldwide Governance Indicators. 




\subsection{Poverty reduction potential of countries and poverty stagnation}

Following the approach of Hausmann and Hidalgo, the available capabilities of a country are given by the capabilities required by the goods in which it has comparative advantage. Thus, to locate a country in the PS, we follow \cite{hidalgo2007product}, and we define its current location at the PS as the set of goods in which it has $\text{ \textbf{RCA}}>1$. From this perspective, we want to explain the stagnation of countries in terms of their location on the PS. Therefore, based on our product poverty measures, the PPI and the Eigenpoverty index, we construct two measures for countries that capture their poverty levels associated with its productive structure. In this order of ideas, we introduce the poverty reduction potential of countries PRP and  the eigenpoverty reduction potential EPRP using both measures, the PPI and the Eigenpoverty index of products, respectively. \\

\begin{itemize}
    \item \textbf{PRP$_c$:} we construct this measure as the average of the values $1-PPI_p$ for every product $p$ in which the country $c$ has revealed comparative advantage, i.e, 
    \begin{eqnarray}
        PRP_c = \frac{\sum_p M_{cp}(1-PPI_p)}{\sum_p M_{cp}}.
    \end{eqnarray}
   In here $M_{cp}$ is 1 if $RCA_{cp}>1$ and $0$ otherwise. This indicator captures the potential that a country has to reduce its current poverty level in terms of its current exports basket.\\
    
    \item \textbf{EPRP$_c$:} analogously, the eigenpoverty reduction potential of a country is defined as: 
    \begin{eqnarray}
        EPRP_c = \text{resc}\left(\frac{\sum_p M_{cp}(1-E_p)}{\sum_p M_{cp}}\right).
    \end{eqnarray}
\end{itemize}

where $\text{resc}:\mathbb{R}^n_+ \to \mathbb{R}^n_+$ is defined as
\begin{eqnarray} \label{eq:resc}
    \text{resc}(x) = \log \left(1+\frac{x}{\min(x>0)}\right).
\end{eqnarray}

Let us notice that the components of the vectors $\text{resc}(x)$ and $x$ have the same ordering and that $\text{resc}(x)_i=0$ if and only if $x_i=0$.\\

Since our ultimate goal is to show that the productive structure of a country explains its future poverty levels and, in particular, a possible stagnation of poverty, we consider the relationship between the current production of a country and its future level of poverty. Thus, since our production data vary between 1995 and 2010, we consider the headcount measure in 2018. We refer the reader to the Appendix (see Section \ref{sec:timeinterval}) for a further discussion about the choosing of the time interval. Actually, for our computations we consider the rescaled headcount measure $RH_c = \text{resc}(H_c)$ in 2018.


\subsection{Robustness and statistical relations} \label{sec:robustness}

In this section, we perform a series of regressions and robustness tests to show the statistical significance of our developed measures as predictors of the stagnation of poverty using information for 92 economies. We start with a model that regresses the headcount measure in 2018 $(RH)$ against the poverty reduction measure (PRP) with other controls. 

\begin{equation}\label{stagnation}
    RH_{2018i} = \alpha_0 + \alpha_1PRP_i + \alpha_lC_{i}+ \varepsilon_i
\end{equation}

where $\alpha_0$ is the constant or intercept of the model, $\alpha_1$ is the coefficient associated with the PRP, $\alpha_l$ is the vector of coefficients associated with the control variables $C$, and $\varepsilon$ is the error term.\\

Since our purpose is to capture stagnation of poverty, the dependent variable in our regression must capture two aspects: high poverty levels and low or negligible variations over time. For instance, the stagnation index of the left hand side of \eqref{stagnation} should be high in the cases of poor countries that have had small improvements in their poverty levels, but also in the case in which countries that evidenced high poverty levels in the first period, that improved slightly in the short run, but whose poverty levels worsen in long periods returning to values close to their original condition. Cases such as countries with low poverty values that have not worsen their situation should correspond to low values of our stagnation index. Mathematically, the following expression represents all the conditions that we need in our independent variable:

\begin{equation} \label{eq:stagnation}
    Stagnation = (1+\Delta HC_{(t,t-8)})HC_{t-8}
\end{equation}

where $\Delta HC_{(t,t-8)}$ is the percent change of the headcount measure between a window of 8 periods and $HC_{t-8}$ is the headcount measure in the period $t-8$. Let us notice that expression \eqref{eq:stagnation} for a percent change close to zero, assigns the level of poverty of period $t-8$, which for non poor economy, it will return a low value (which may be close to zero) and for a poor economy, it will assign a higher value. If the percent change was positive, whether the economy is non poor or poor, the stagnation will have a greater value of poverty in $t-8$ for each economy, and if the percent change is negative, the stagnation will return a lower value of the poverty in $t-8$. Now, computing for \eqref{eq:stagnation} we get

\begin{eqnarray} 
    Stagnation =&  (1+\frac{(HC_t-HC_{t-8})}{HC_{t-8}})HC_{t-8}\\
    =&  HC_{t-8}+(HC_t-HC_{t-8}) = HC_t. \label{eq:stagnationsolved}
\end{eqnarray}

If we replace the headcount measures of 2010 and 2018 in equation \eqref{eq:stagnationsolved}, we get that stagnation is the headcount measure in 2018.\\

In Table \ref{table:reg1}, we present various models which compare the poverty in 2018 $(RH)$ against the poverty reduction measure (PRP), log of GDP per capita at constant PPP US\$ prices (log(GDP)), years of schooling, log of the population (log(population)), economic complexity (ECI) of \cite{hidalgo2009building}, and  institutional variables such as government effectiveness, political stability, voice and accountability, regulatory quality, and rule of low. As mentioned before, the rescaled poverty is computed with data of 2018, whereas the PRP was calculated using an average of the $M_{cp}$ matrix and PPI for 1995-2010. For the rest of variables, we used the average between 2006 and 2010.\\

\begin{table}[h!] \centering
\resizebox{\columnwidth}{!}{%
\begin{tabular}{@{\extracolsep{5pt}}lcccccc}
\\[-1.8ex]\hline
\hline \\ [-1.8ex]
& \multicolumn{6}{c}{\textit{Dependent variable: rescaled headcount 2018 ($RH$)}} \
\cr \hline\\[-1.8ex]
\\[-1.8ex] & (1) & (2) & (3) & (4) & (5) & (6) \\
\hline \\[-1.8ex]
 Constant & 29.772$^{***}$ & 7.131$^{**}$ & 30.005$^{***}$ & 29.837$^{***}$ & 30.519$^{***}$ & 26.798$^{***}$ \\
  & (7.763) & (3.321) & (8.015) & (7.707) & (7.894) & (3.895) \\
 PRP & -27.892$^{***}$ & & -30.388$^{***}$ & -28.351$^{***}$ & -26.372$^{***}$ & -21.795$^{***}$ \\
  & (8.741) & & (8.967) & (8.250) & (8.865) & (5.729) \\
 years of schooling  & -0.232$^{**}$ & -0.265$^{***}$ & & -0.235$^{**}$ & -0.251$^{***}$ & -0.247$^{***}$ \\
  & (0.092) & (0.096) & & (0.089) & (0.093) & (0.089) \\
 log(GDP pc) & -0.044$^{}$ & -0.305$^{}$ & -0.175$^{}$ & & -0.226$^{}$ & -0.115$^{}$ \\
  & (0.260) & (0.261) & (0.263) & & (0.248) & (0.193) \\
 Government effectiveness & -1.627$^{**}$ & -1.401$^{}$ & -1.840$^{**}$ & -1.675$^{**}$ & & \\
  & (0.815) & (0.856) & (0.837) & (0.759) & & \\
 ECI & 0.468$^{}$ & -0.834$^{**}$ & 0.217$^{}$ & 0.480$^{}$ & 0.264$^{}$ & \\
  & (0.546) & (0.383) & (0.554) & (0.538) & (0.546) & \\
 log(population) & 0.094$^{}$ & 0.162$^{}$ & 0.161$^{}$ & 0.095$^{}$ & 0.059$^{}$ & \\
  & (0.136) & (0.142) & (0.138) & (0.135) & (0.137) & \\
 Political stability  & -0.285$^{}$ & -0.136$^{}$ & -0.358$^{}$ & -0.287$^{}$ & -0.224$^{}$ & \\
  & (0.294) & (0.306) & (0.302) & (0.292) & (0.297) & \\
 Voice and accountability & 0.696$^{**}$ & 0.795$^{**}$ & 0.723$^{**}$ & 0.689$^{**}$ & 0.806$^{**}$ & \\
  & (0.323) & (0.339) & (0.333) & (0.318) & (0.324) & \\
 Regulatory quality & 0.479$^{}$ & 0.385$^{}$ & 0.339$^{}$ & 0.480$^{}$ & 0.174$^{}$ & \\
  & (0.568) & (0.599) & (0.584) & (0.565) & (0.557) & \\
 Rule of low & 0.511$^{}$ & 0.589$^{}$ & 0.850$^{}$ & 0.519$^{}$ & -0.500$^{}$ & \\
  & (0.708) & (0.746) & (0.717) & (0.702) & (0.503) & \\
\hline \\[-1.8ex]
 Observations & 92 & 92 & 92 & 92 & 92 & 92 \\
 $R^2$ & 0.681 & 0.641 & 0.655 & 0.681 & 0.665 & 0.628 \\
 Adjusted $R^2$ & 0.641 & 0.601 & 0.618 & 0.646 & 0.628 & 0.615 \\
 Residual Std. Error & 1.429(df = 81) & 1.507(df = 82) & 1.476(df = 82) & 1.421(df = 82) & 1.455(df = 82) & 1.480(df = 88)  \\
 F Statistic & 17.268$^{***}$ & 16.237$^{***}$ & 17.330$^{***}$ & 19.413$^{***}$  & 18.085$^{***}$  & 49.508$^{***}$  \\
  &  (df = 10.0; 81.0) & (df = 9.0; 82.0) & (df = 9.0; 82.0) &  (df = 9.0; 82.0) & (df = 9.0; 82.0) & (df = 3.0; 88.0) \\
\hline
\hline \\[-1.8ex]
\textit{Note:} & \multicolumn{6}{r}{$^{*}$p$<$0.1; $^{**}$p$<$0.05; $^{***}$p$<$0.01} \\
\end{tabular}%
}
\caption{Cross-section OLS regression models using PRP}
\label{table:reg1}
\end{table}

Model 1 shows that all variables contribute to explaining 64\% of the model's total variance. Furthermore, the PRP  consistently explains future poverty in all models with a negative and statistically significant relationship. This means that countries with a particular export basket with a higher poverty reduction potential will tend to reduce their future poverty levels. Interestingly, ECI was not statistically significant in any model but in model 2, where we dropped the PRP. The latter suggests that the introduction of poverty in the PRP captures aspects of capabilities that are not explained by the sophistication of the products. On the other hand, years of schooling are also a consistent predictor of stagnation with a negative and significant coefficient. Additionally, variables that are constantly related to poverty, such GDP per capita or log of population, were not statistically significant. However, some institutional variables such as government effectiveness and voice and accountability appeared with a significant correlation.\\


\begin{figure}[h!]
    \centering
    \includegraphics[scale=0.45]{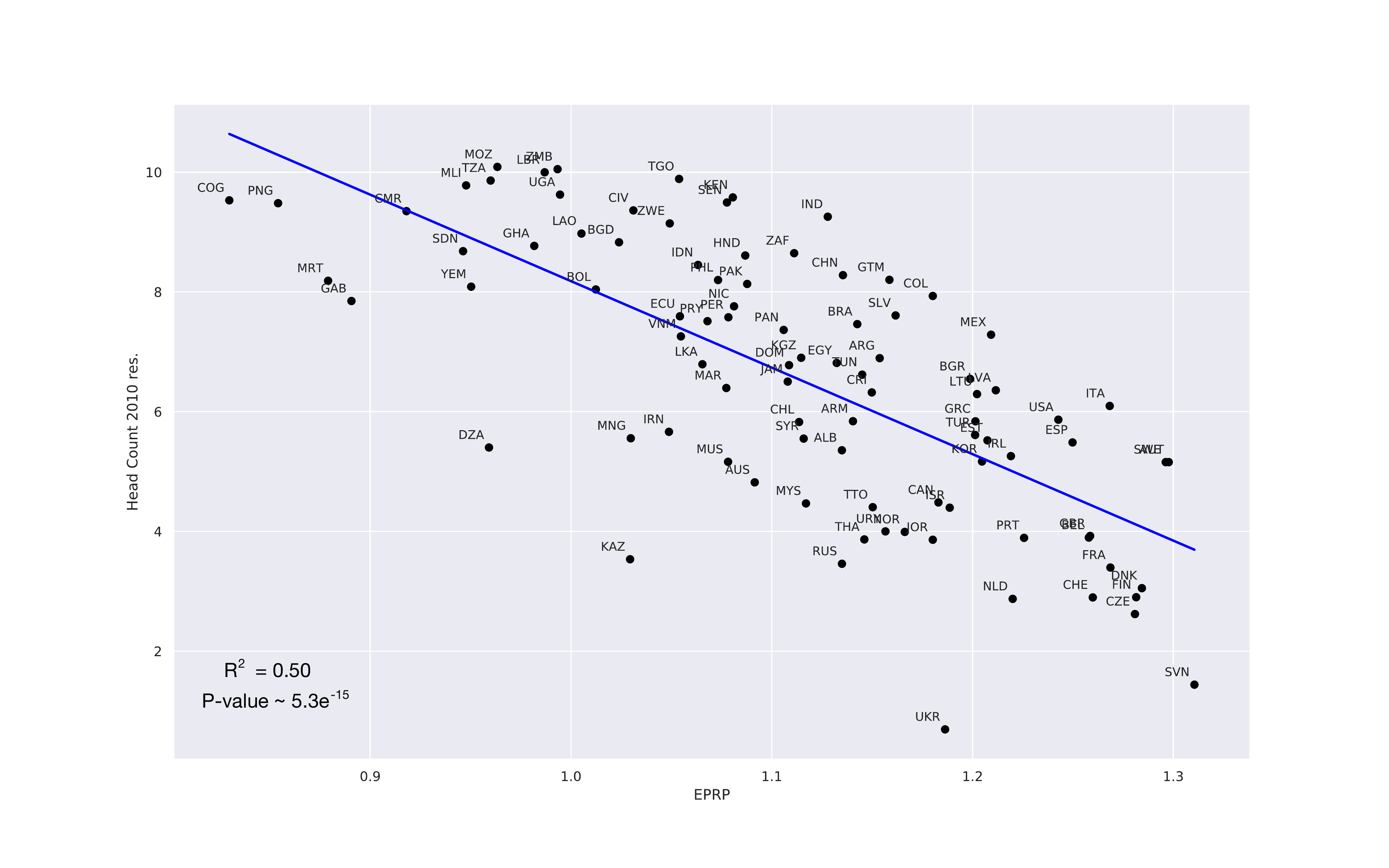}
    \caption{Rescaled headcount measure in 2010 versus EPRP.}
    \label{fig:reg3}
\end{figure}

The semi-partial correlation of PRP (the difference in adjusted R$^2$ when only the PRP is removed from the model that contemplates all variables) is 4\%, being the variable that contributes the most to the total variance explanation of the model. Analogously, the semi-partial correlation of years of schooling is 2.3\%. In contrast, the log of GDP does not contribute to the model, while institutional variables and ECI contributes in 2.6\%. \\

We perform a similar analysis for the case of the EPRP. Given the construction of Eigenpoverty as a long-run measure of poverty, we carry out an analysis in two stages in the same spirit of \cite{hausmann2021implied}. First, we make a regression  of the rescaled headcount index in 2010, using equation \eqref{eq:resc}, against the EPRP. This is 

\begin{equation} \label{eq:regstage1}
     RH_{2010i} = \alpha_0 + \alpha_1EPRP_i + \varepsilon_i
\end{equation}

This regression is statistically significant with a negative coefficient, as depicted in Figure \ref{fig:reg3}. Then, we extract the residuals of this regression (the real values of $RH_{2010i}$ minus the predicted in equation \eqref{eq:regstage1}), as

\begin{equation}
     r_i = RH_{2010i}-\hat{RH}_{2010i},
\end{equation}

and we use them to explain the future values of the rescaled headcount measure (in 2018) 

\begin{equation}
     RH_{2018i} = \alpha_0 + \alpha_1r_i + \alpha_lC_i + \varepsilon_i.
\end{equation}

The rationale behind of this two-stage regression consists in showing that the difference between the current poverty and the poverty explained by the Eigenpoverty (i.e., the residuals of the first regression) have explanatory power over the future levels of poverty. More precisely, if a country has more poverty than the predicted by its Eigenpoverty (its associated residual is positive), it is expected to decrease in the future. Conversely, if the actual level of poverty of a country lies below its predicted level of poverty, then it is expected that the poverty levels of this country adjust in the future by an increment. This intuition is confirmed by a series of regressions summarized in Table \ref{table:reg3}. This table illustrates five models that regress the future rescaled headcount measure ($RH$) against the residuals of the first regression (see Figure \ref{fig:reg3}), adding systematically all the other control variables utilized in Table \ref{table:reg1} for the sake of robustness. The residuals are statistically significant in all models with a negative relationship, as expected, which means, graphically, that those countries that are far below or far above the prediction line in Figure \ref{fig:reg3} will increase or decrease its expected future poverty levels, respectively. Comparing the two full models of both tables (column 1 of \ref{table:reg1} and column 2 of Table
\ref{table:reg3}), we can see that including the residuals in the model increases the explanation of the total variance by about 18.5\%, being the most important predictor. Years of schooling becomes statistically irrelevant in the presence of residuals. It means that residuals of the EPRP can largely explain and predict stagnation, recollecting information of years of schooling and institutional variables that were important before  adding this variable. 


\begin{table}[h!] \centering
\scalebox{0.683}{
\begin{tabular}{@{\extracolsep{5pt}}lccccc}
\\[-1.8ex]\hline
\hline \\[-1.8ex]
& \multicolumn{5}{c}{\textit{Dependent variable: rescaled headcount 2018 ($RH$)}} \
\cr \hline\\[-1.8ex]
\\[-1.8ex] & (1) & (2) & (3) & (4) & (5) \\
\hline \\[-1.8ex]
 Constant & 5.255$^{***}$ & 25.599$^{***}$ & 7.131$^{**}$ & 5.120$^{**}$ & 6.265$^{***}$ \\
  & (0.197) & (5.426) & (3.321) & (2.323) & (2.305) \\
 Resid. & -0.925$^{***}$ & -0.777$^{***}$ & & -0.803$^{***}$ & -0.807$^{***}$ \\
  & (0.125) & (0.083) & & (0.084) & (0.091) \\
 PRP & & -24.990$^{***}$ & & & \\
  & & (6.097) & & & \\
 years of schooling & & 0.005$^{}$ & -0.265$^{***}$ & & -0.024$^{}$ \\
  & & (0.069) & (0.096) & & (0.075) \\
 log(GDP pc) & & 0.215$^{}$ & -0.305$^{}$ & -0.021$^{}$ & -0.095$^{}$ \\
  & & (0.183) & (0.261) & (0.188) & (0.182) \\
 Government effectiveness & & -1.057$^{*}$ & -1.401$^{}$ & -0.850$^{}$ & -1.614$^{***}$ \\
  & & (0.571) & (0.856) & (0.617) & (0.283) \\
 ECI & & -0.353$^{}$ & -0.834$^{**}$ & -1.569$^{***}$ & \\
  & & (0.390) & (0.383) & (0.254) & \\
 log(population) & & -0.033$^{}$ & 0.162$^{}$ & 0.029$^{}$ & 0.003$^{}$ \\
  & & (0.096) & (0.142) & (0.102) & (0.103) \\
 Political stability & & -0.134$^{}$ & -0.136$^{}$ & -0.001$^{}$ & 0.032$^{}$ \\
  & & (0.205) & (0.306) & (0.220) & (0.221) \\
 Voice and accountability & & 0.191$^{}$ & 0.795$^{**}$ & 0.265$^{}$ & 0.314$^{}$ \\
  & & (0.231) & (0.339) & (0.250) & (0.251) \\
 Regulatory quality & & -0.380$^{}$ & 0.385$^{}$ & -0.504$^{}$ & -0.652$^{}$ \\
  & & (0.407) & (0.599) & (0.434) & (0.428) \\
 Rule of low & & 0.972$^{*}$ & 0.589$^{}$ & 1.083$^{**}$ & 0.536$^{}$ \\
  & & (0.495) & (0.746) & (0.525) & (0.386) \\
\hline \\[-1.8ex]
 Observations & 92 & 92 & 92 & 92 & 92 \\
 $R^2$ & 0.378 & 0.847 & 0.641 & 0.815 & 0.811 \\
 Adjusted $R^2$ & 0.372 & 0.826 & 0.601 & 0.794 & 0.790 \\
 Residual Std. Error & 1.892(df = 90) & 0.996(df = 80) & 1.507(df = 82) & 1.082(df = 82) & 1.094(df = 82)  \\
 F Statistic & 54.790$^{***}$ & 40.252$^{***}$ & 16.237$^{***}$ & 40.052$^{***}$ & 39.002$^{***}$  \\
  & (df = 1.0; 90.0) &  (df = 11.0; 80.0) &  (df = 9.0; 82.0) &  (df = 9.0; 82.0) & (df = 9.0; 82.0) \\
\hline
\hline \\[-1.8ex]
\textit{Note:} & \multicolumn{5}{r}{$^{*}$p$<$0.1; $^{**}$p$<$0.05; $^{***}$p$<$0.01} \\
\end{tabular}%
}
\caption{Cross-section OLS regression models using residuals of Eigenpoverty}
\label{table:reg3}
\end{table}

\section{Case study}
\label{sec:caseofstudy}
\begin{figure}[h!] 
\begin{multicols}{3}
    \includegraphics[width=\linewidth]{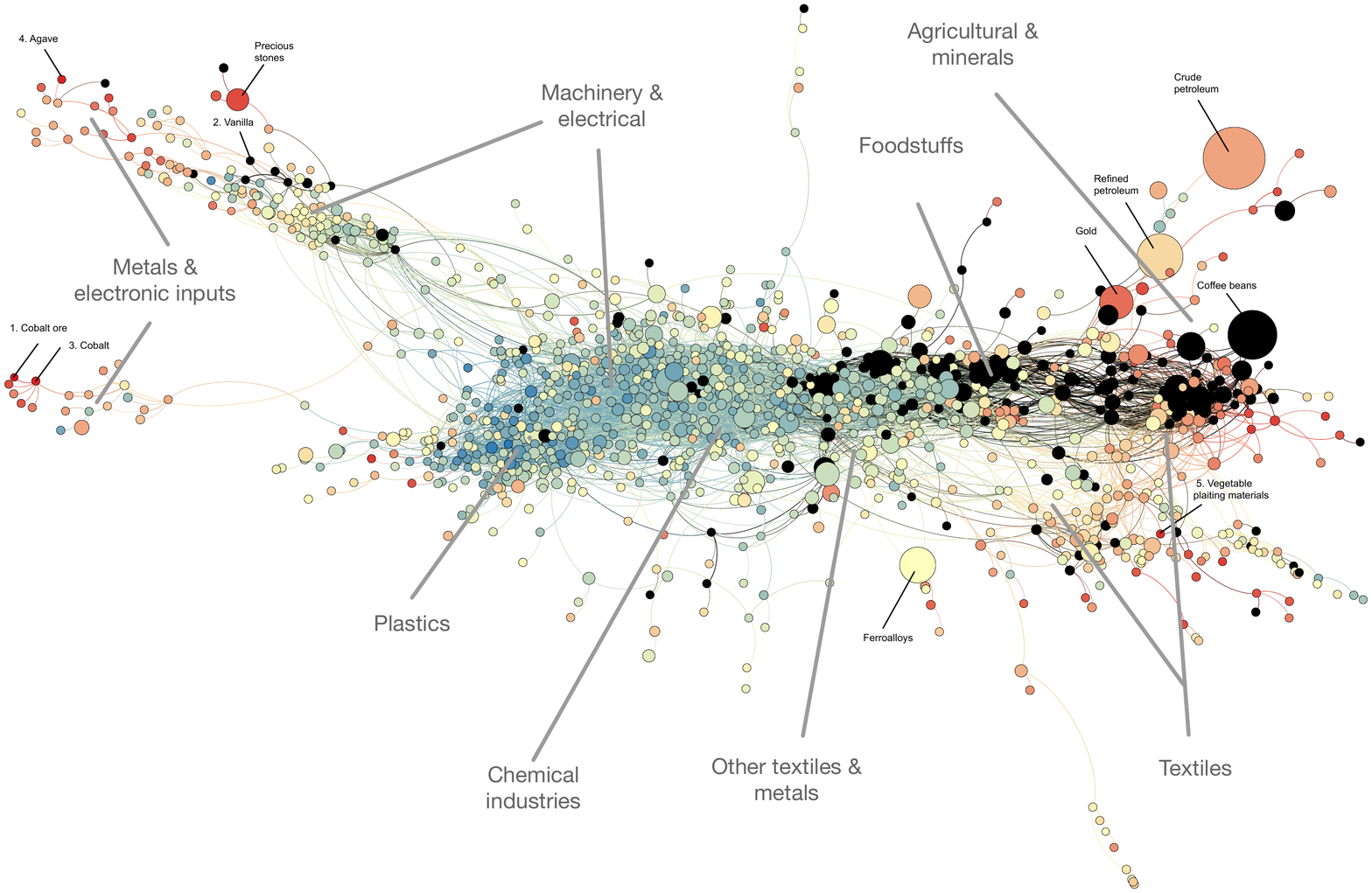}\par\caption{Costa Rica 1995} \label{fig:cri95}
    \includegraphics[width=\linewidth]{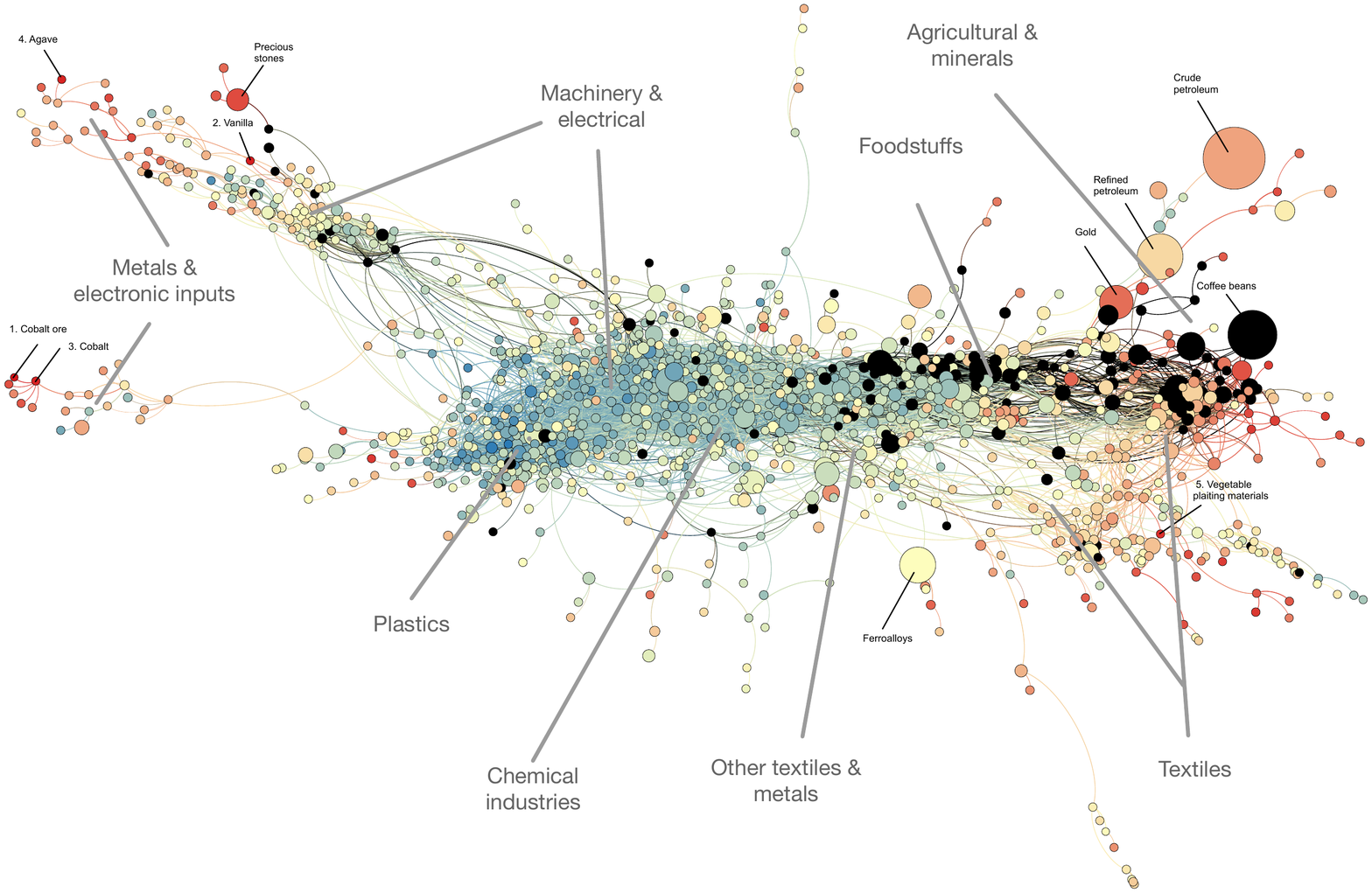}\par\caption{Costa Rica 2003} \label{fig:cri03}
    \includegraphics[width=\linewidth]{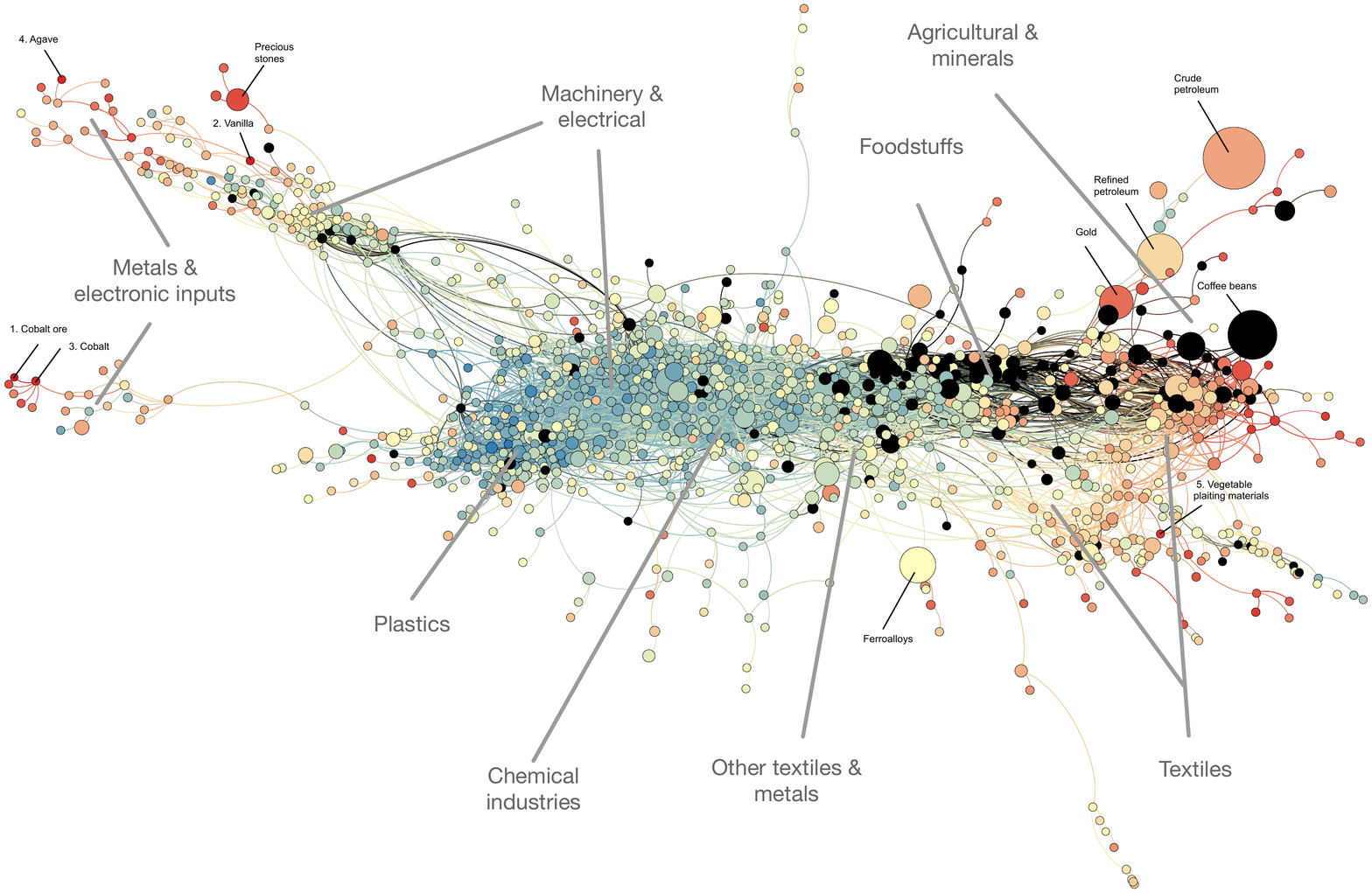}\par\caption{Costa Rica 2010} \label{fig:cri10}
\end{multicols}
\begin{multicols}{3}
    \includegraphics[width=\linewidth]{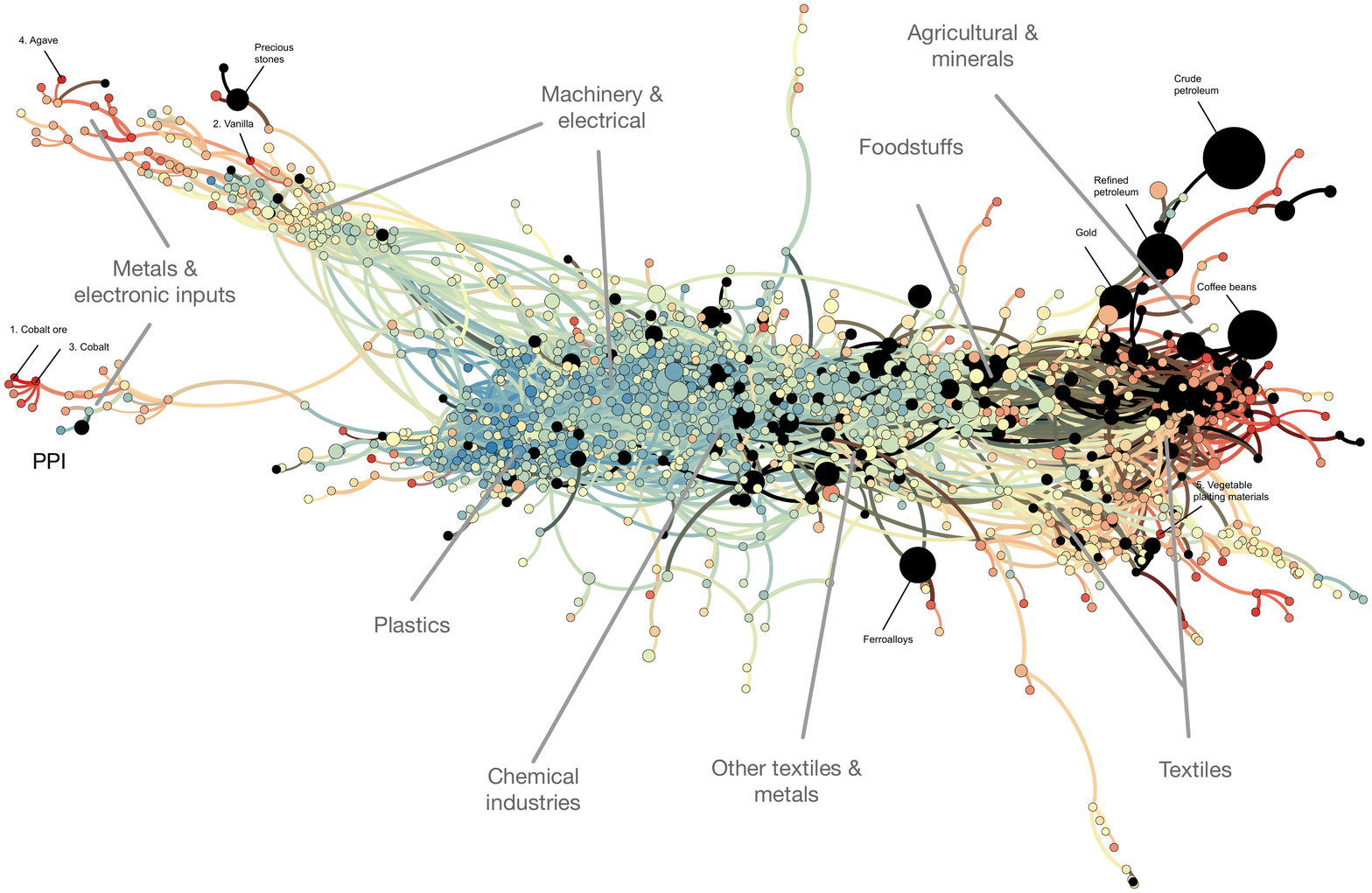}\par\caption{Colombia 1995}\label{fig:col95}
    \includegraphics[width=\linewidth]{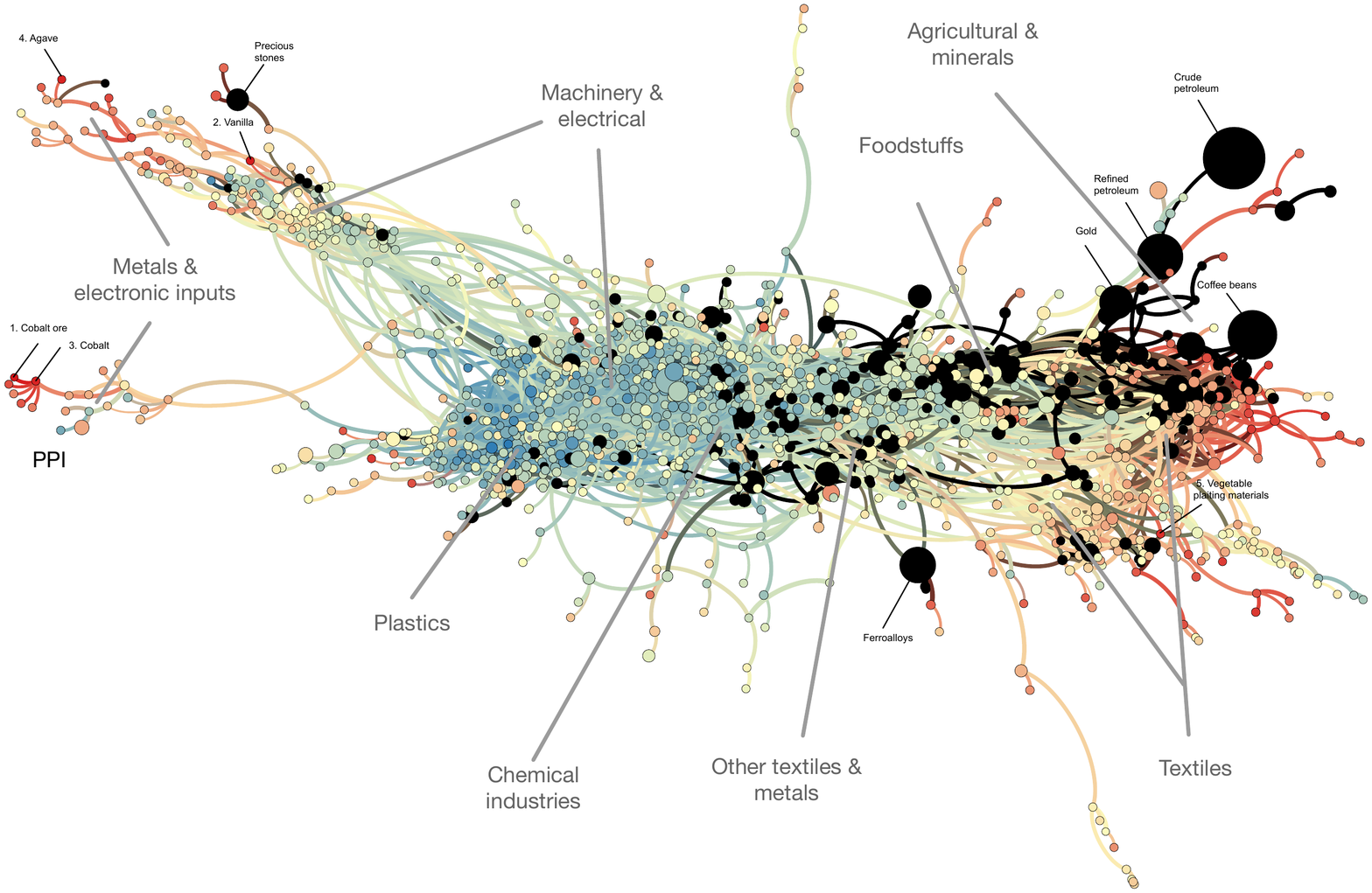}\par\caption{Colombia 2003}\label{fig:col03}
    \includegraphics[width=\linewidth]{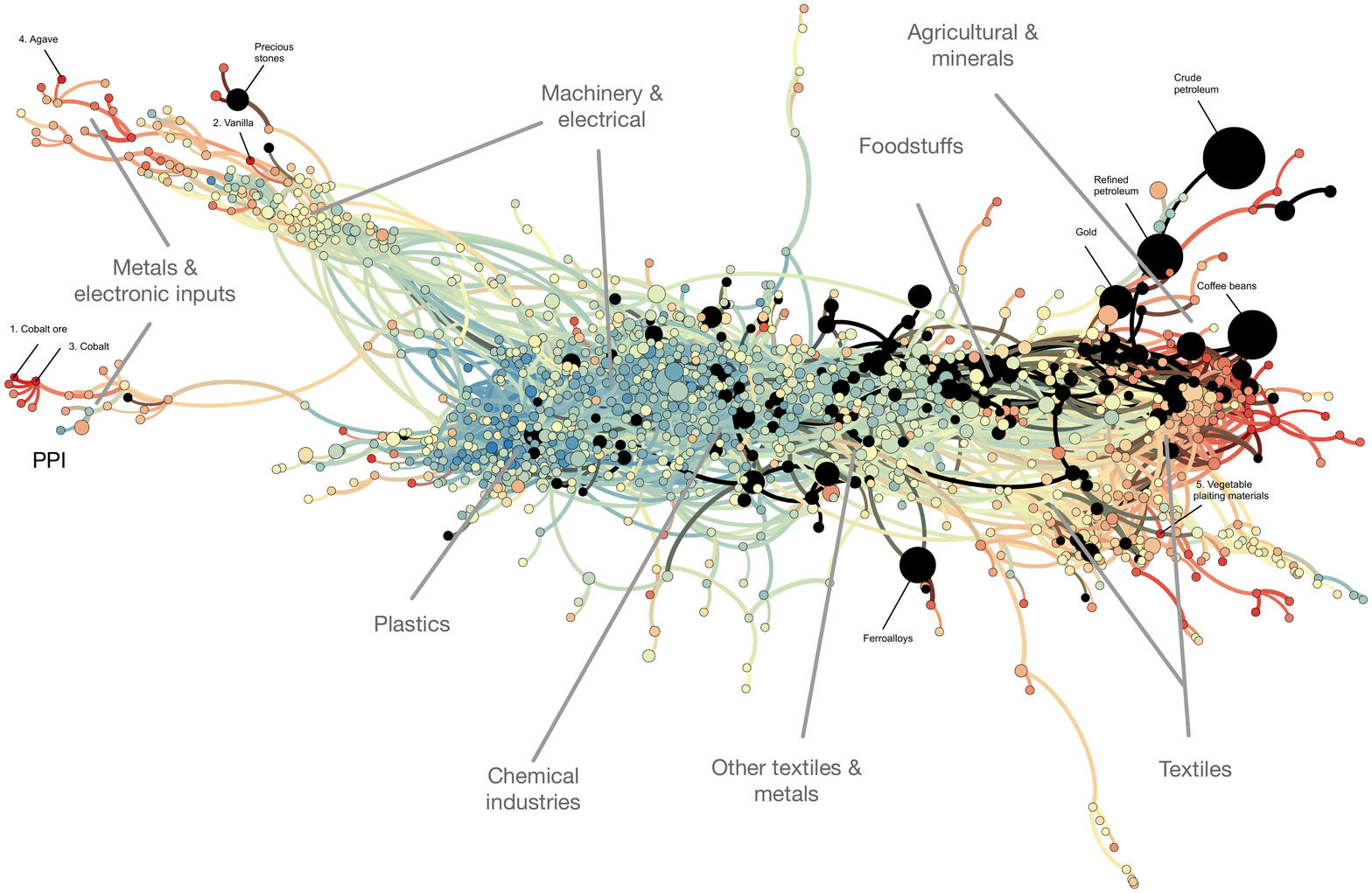}\par\caption{Colombia 2010}\label{fig:col10}
\end{multicols}
\begin{multicols}{3}
    \includegraphics[width=\linewidth]{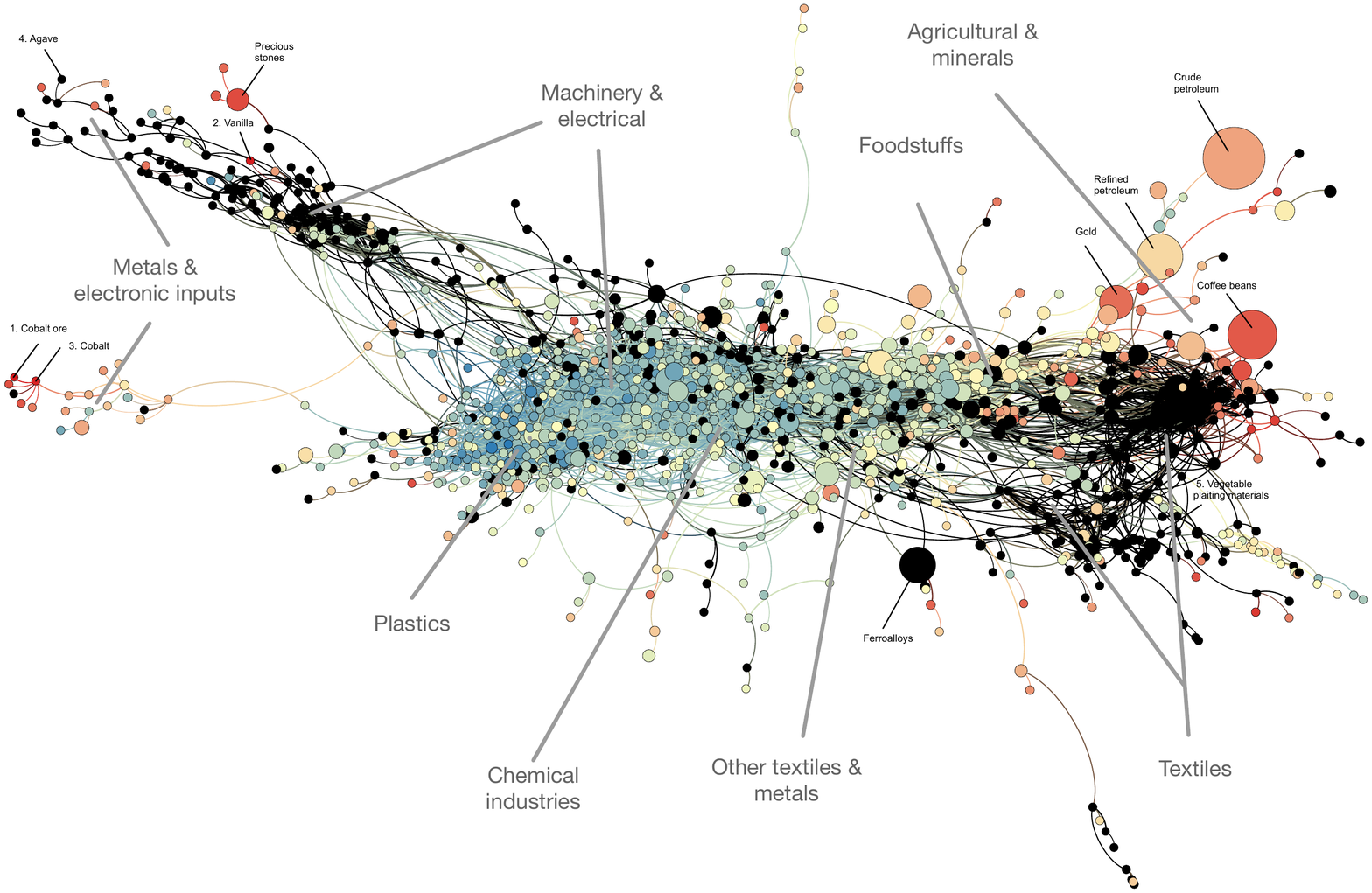}\par\caption{China 1995}\label{fig:chn95}
    \includegraphics[width=\linewidth]{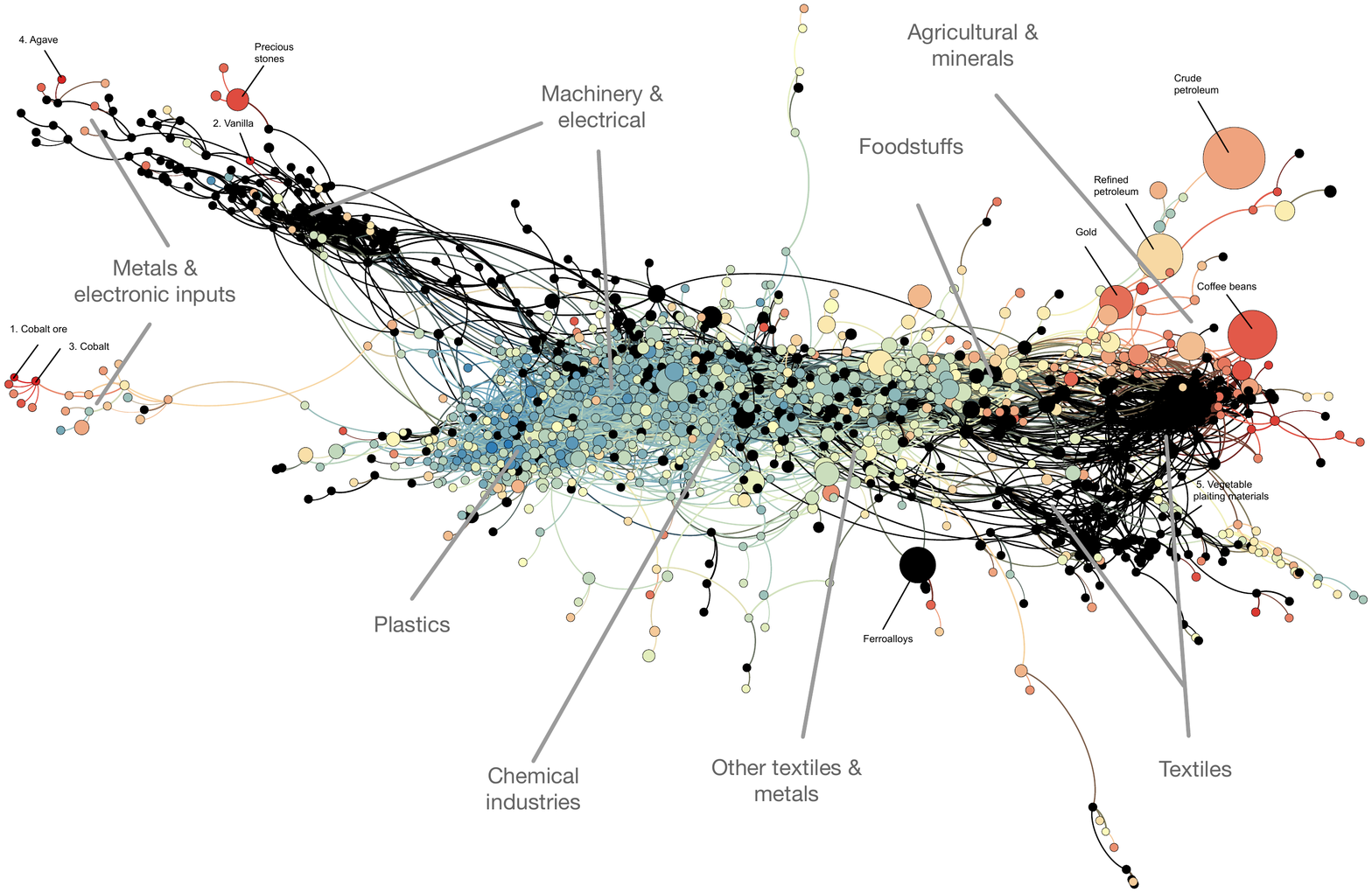}\par\caption{China 2003}\label{fig:chn03}
    \includegraphics[width=\linewidth]{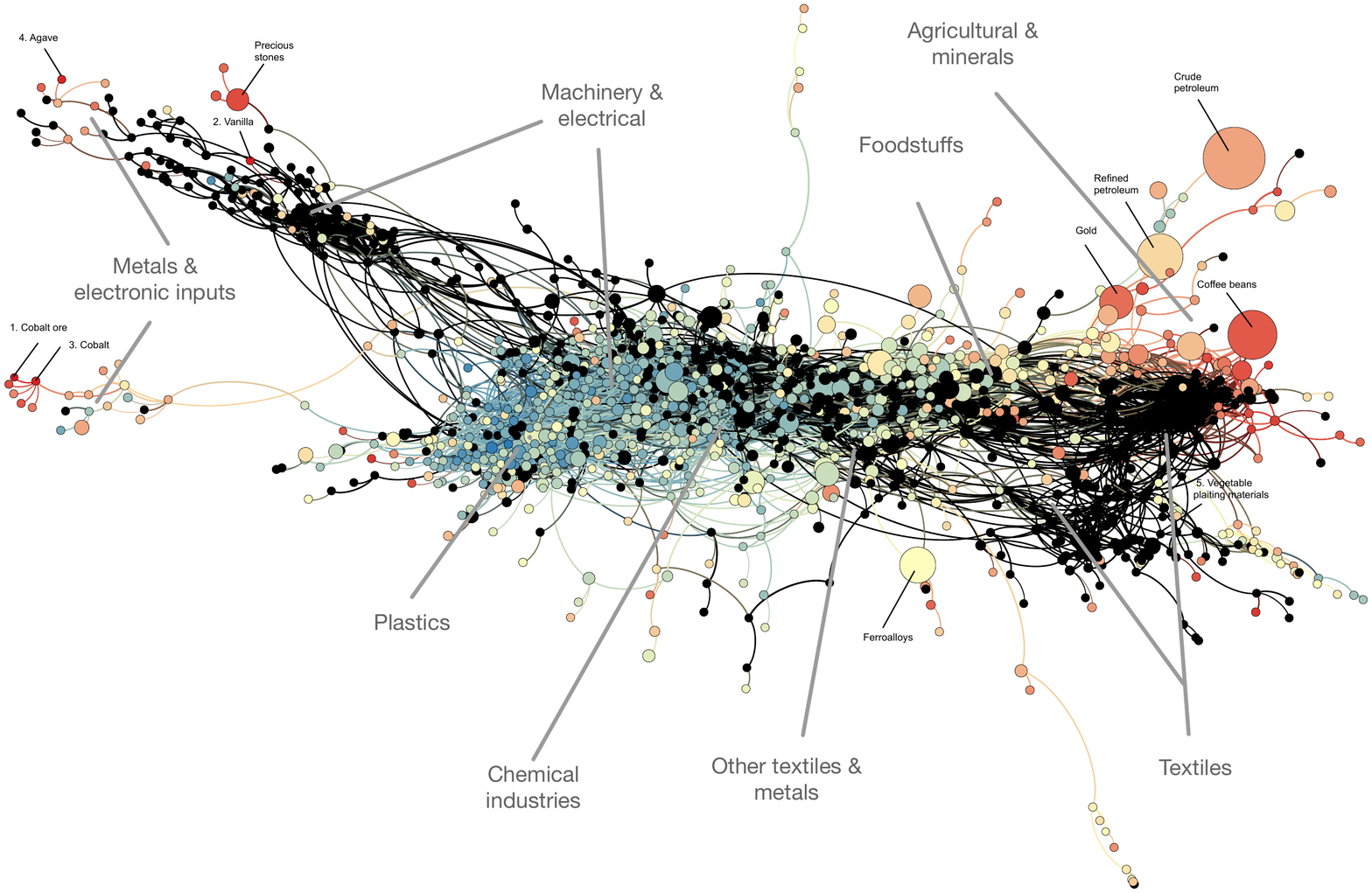}\par\caption{China 2010} \label{fig:chn10}
\end{multicols}
\caption{\small This collection of figures depicts the temporal changes on the PS of three economies, colored by the distribution of the PPI. Figures \ref{fig:cri95}, and \ref{fig:cri03}, \ref{fig:cri10} represent the location on the PS for Costa Rica in 1995, 2003, and 2010, respectively. Analogously, Figures \ref{fig:col95}, \ref{fig:col03}, and \ref{fig:col10} are the location on the PS for Colombia in 1995, 2003, 2010. Finally, Figures \ref{fig:chn95}, \ref{fig:chn03}, and \ref{fig:chn10} represent the same procedure of Costa Rica and and Colombia, but for China. The black spots in a PS indicate that the country exported those products with RCA>1.}
\label{fig:collection}
\end{figure}

This section presents three cases that aim to provide evidence of the relation between the location on the PS and poverty traps. First, we present the cases of three economies: Colombia, Costa Rica and China. The attempt of this comparison is to show how our metrics enable to capture substantial differences in poverty improvement (China vs Costa Rica and Colombia) or more nuanced differences (Costa Rica vs. Colombia). Second, for this analysis, we locate all countries on the PS with the PPI distribution taking three years to compare: 1995, 2003, and 2010. This is represented in the collection of figures in Figure \ref{fig:collection}. \\

\begin{figure}[h!]
    \centering
    \includegraphics[scale=0.4]{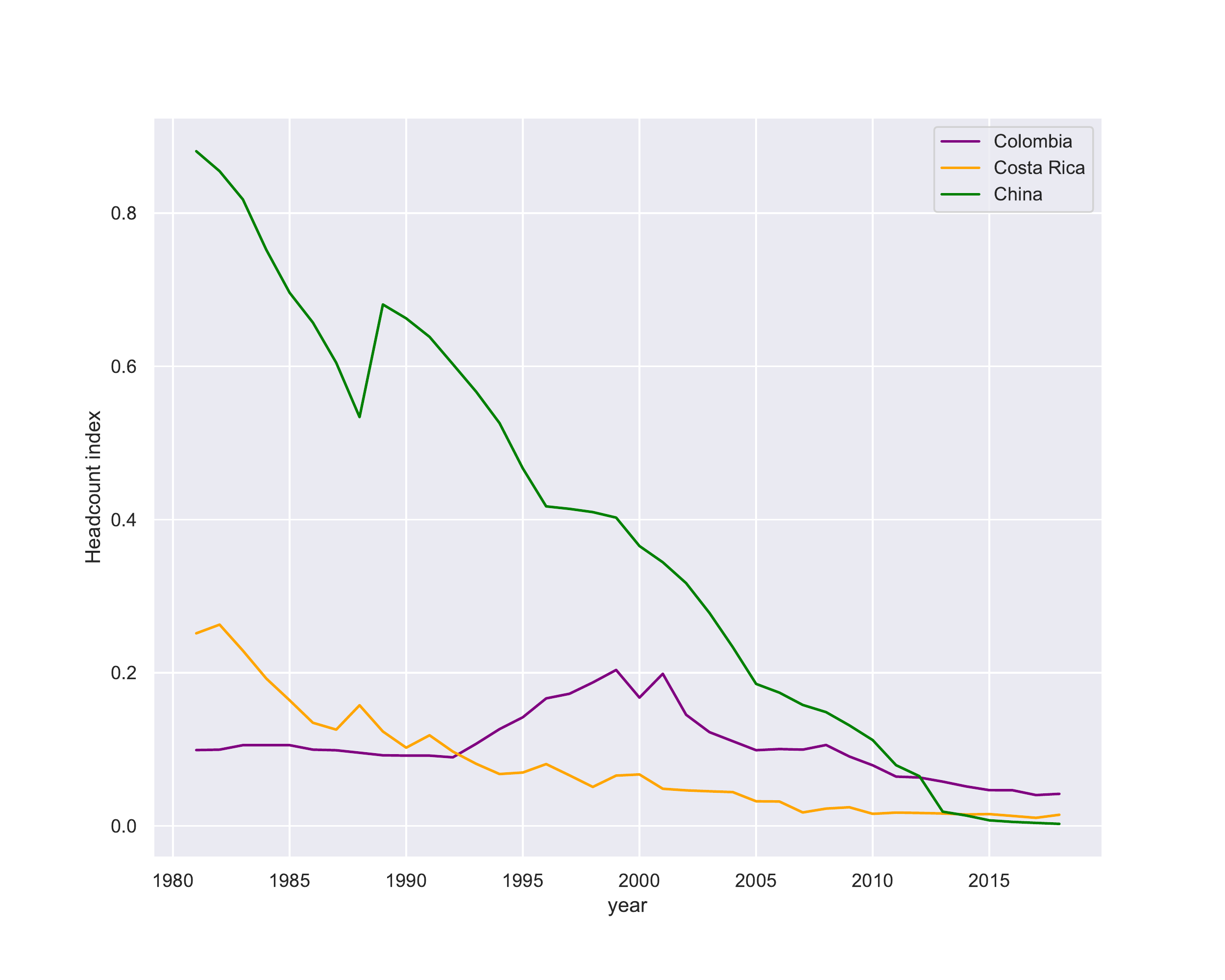}
    \caption{Headcount measure of Colombia, Costa Rica, and China for 1980-2018}
    \label{fig:headcount_three_countries}
\end{figure}

 We start comparing Costa Rica and Colombia. It is remarkable that even if Colombia has been historically more diverse than Costa Rica as we can see in Figure \ref{fig:collection}  (for instance, they exported 144 and 130 products with \textbf{RCA>1} in 2010, respectively), it has also been poorer than Costa Rica (see Figure \ref{fig:headcount_three_countries}. Costa Rica and Colombia have almost kept their production in the same zones on the PS during 15 years, i.e., Costa Rica has been producing goods of agricultural and mineral nature, some textiles related to natural resources, and foodstuffs. Costa Rica lost advantage in some products, mainly in textiles, but in 2010 it gained more in foodstuffs and chemicals associated with the existing industries. During the three periods, Colombia was located in textiles related to natural resources, foodstuffs, agricultural, and mineral products, but also in chemicals and plastics. Comparing the location of Colombia in the PS in 1995 to its PS in 2003, the country lost importance in the production of some goods since it abandoned products in the periphery, primarily agricultural products, and it gained products in electronics and in the center of the PS. Despite this, in 2010 Colombia lost importance in goods related to electronics and foodstuffs. It is noticeable that Colombia and Costa Rica have products, in their majority, in red, dark orange,  and light orange, and in their minority, in yellow and green zones. \\
 
 On the other hand, we have the case of China, which is an economy that started with a high poverty rate and has dramatically decreased its poverty levels. In 1995 China was located in machinery, electronics, and textiles. In 2003 it industrialized more to the center of the PS towards more electronics. Finally, in 2010, it gained more products towards the center of the PS within the industries that it already had and within chemicals. Let us notice that China is  located on the PS in products in light orange, yellow, green, and blue.\\

\begin{figure}[h!]
    \centering
    \begin{subfigure}[h]{0.49\textwidth}
    \centering
        \includegraphics[scale=0.37]{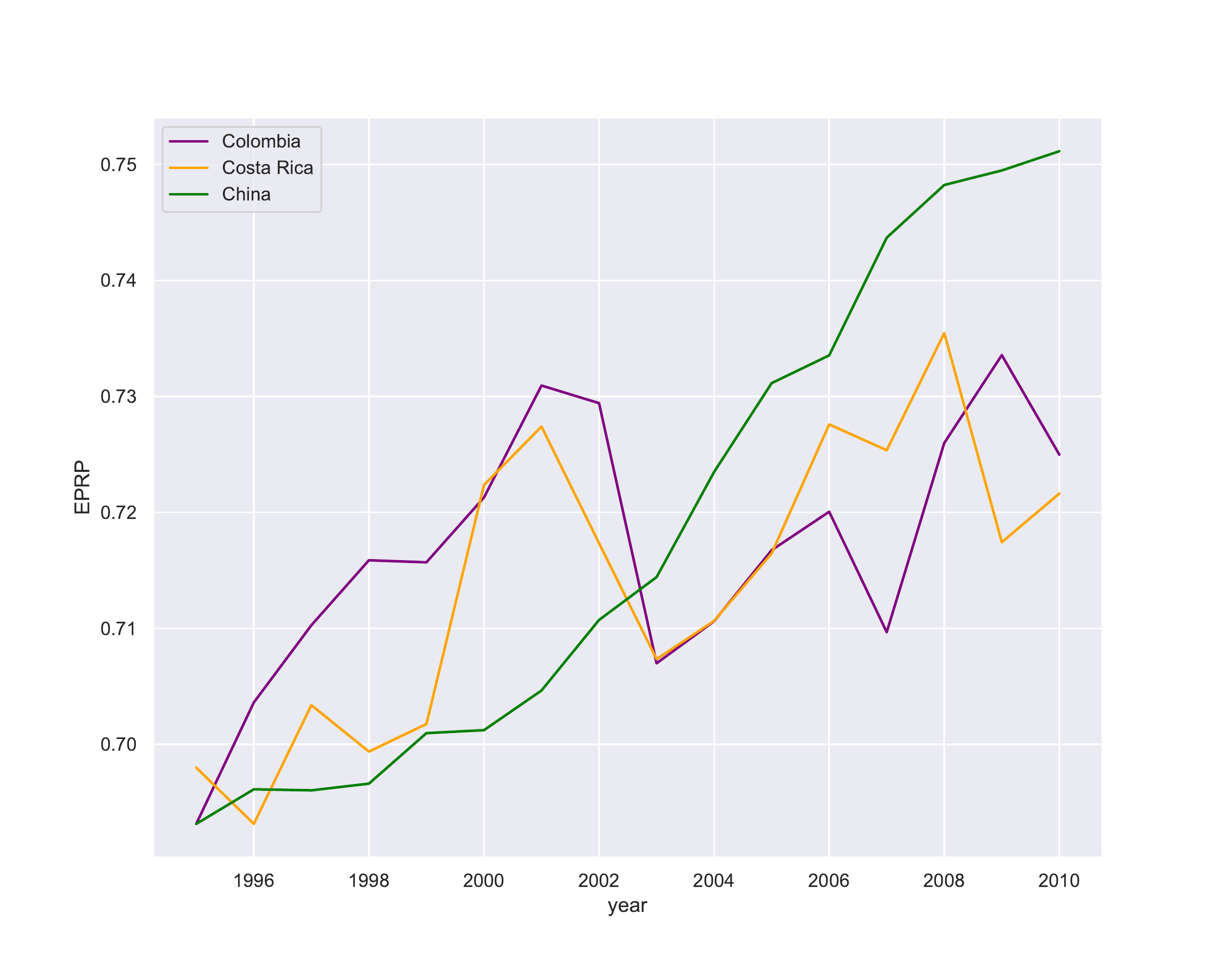}
        \caption{EPRP over time for Colombia, Costa Rica, and China}
        \label{fig:EPRP_three_countries}
    \end{subfigure}
    \hfill
    \begin{subfigure}[h]{0.49\textwidth}
    \centering
        \includegraphics[scale=0.37]{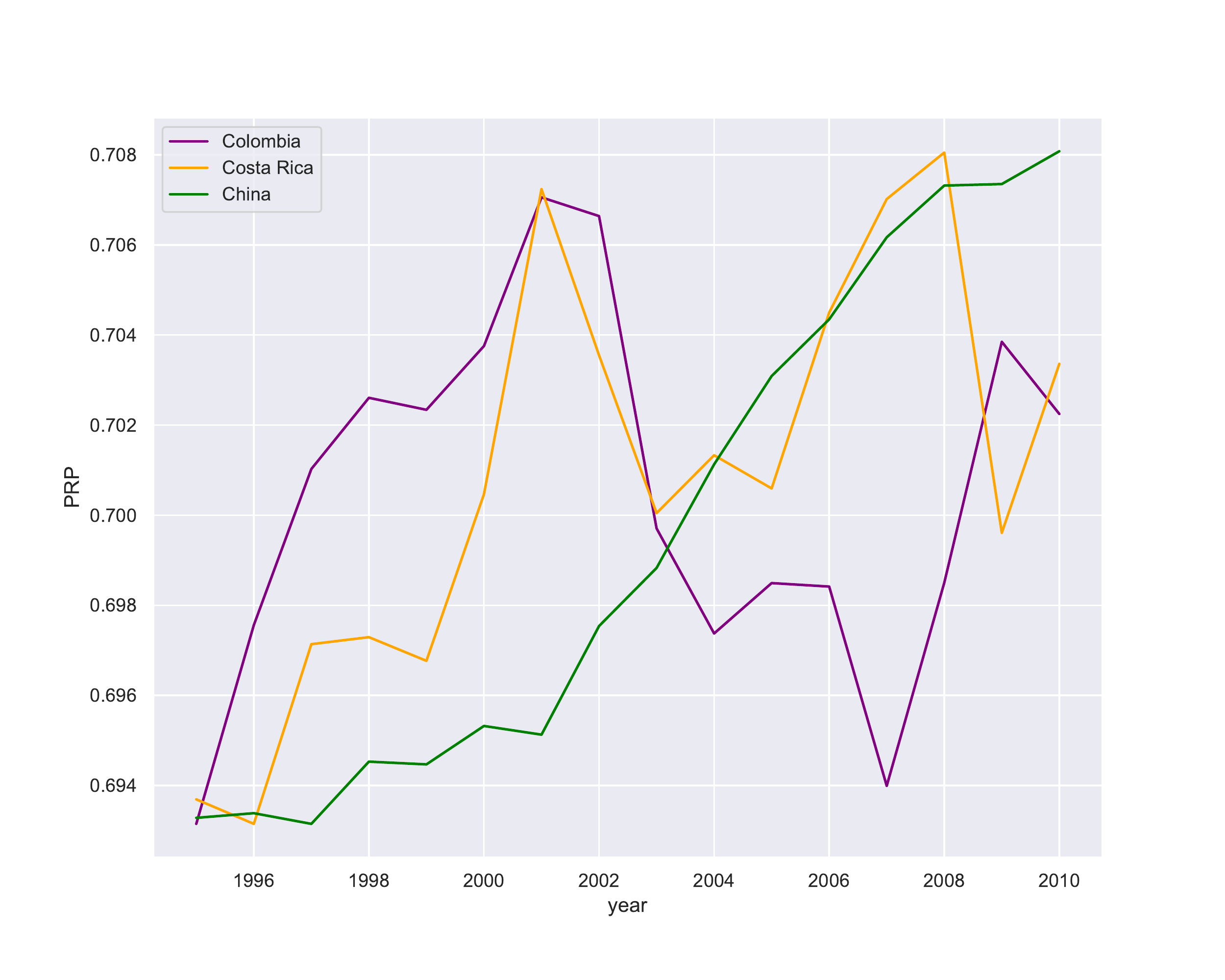}
        \caption{PRP over time for Colombia, Costa Rica, and China}
        \label{fig:PRP_three_countries}
    \end{subfigure}
    \caption{Comparison between three economies for EPRP and PRP}
    \label{fig:three_countries_EPRP_PRP}
\end{figure}

We proceed now to discuss the evolution of the production depicted in Figure \ref{fig:collection} in light of the changes of the EPRP and PRP for the three economies. Figure \ref{fig:three_countries_EPRP_PRP} exhibits the possibilities of improving the current poverty levels of Costa Rica, Colombia, and China during the period 1995-2010 through the EPRP and the PRP. In both Figures (\ref{fig:EPRP_three_countries} and \ref{fig:PRP_three_countries}) China has a consistent and sustained increasing potential of reducing its poverty. On the other hand, the values of the indices for Colombia and Costa Rica have oscillated, although they have presented similar values of EPRP  as shown in Figure \ref{fig:EPRP_three_countries}. However, they have diverged in certain extent in the case of the PRP, as depicted in Figure \ref{fig:PRP_three_countries}. The oscillation of these patterns are related to the loss and gain of the advantage in the production commented above. Therefore, it is consequent to the definition of the PRP and EPRP metrics as measures of the  possibilities in the short-run and long-run to mitigate extreme poverty, respectively. \\

\begin{figure}[h!]
    \centering
    \includegraphics[trim={0cm 3cm 0cm 4cm},clip, scale=0.6]{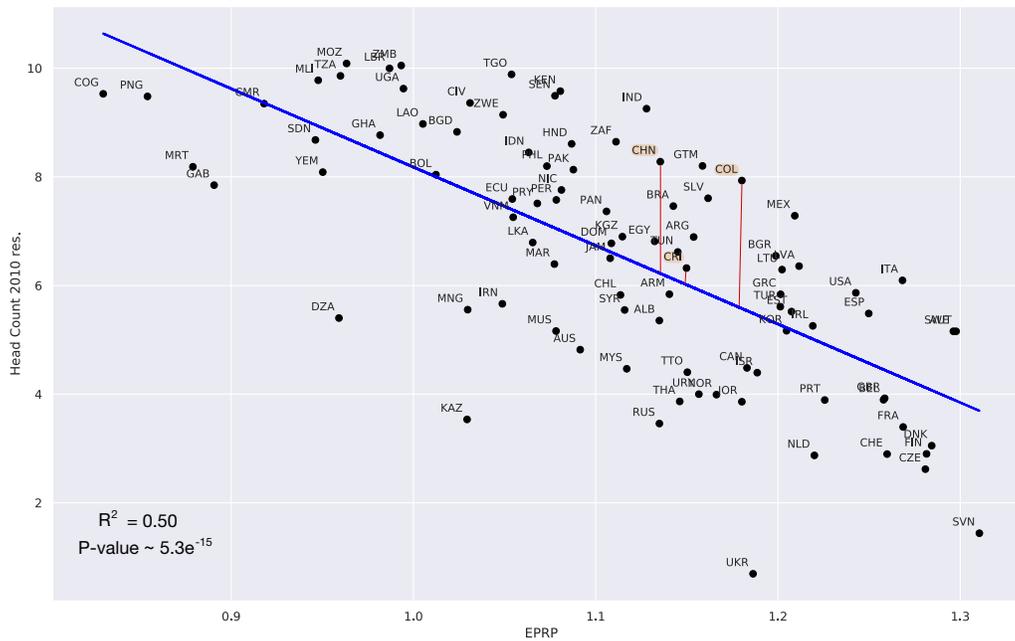}
    \caption{Prediction of the rescaled headcount measure of Costa Rica, Colombia, and China according to their levels of long run potential to reduce poverty (EPRP).}
    \label{fig:hc_vs_EPRP_high}
\end{figure}

In Figure \ref{fig:hc_vs_EPRP_high}, we depicted the distance of Costa Rica, Colombia, and China to their expected values of poverty according to their productive structure. Based on this chart, we find that Colombia should have similar levels of poverty more like Costa Rica, whereas this latter country is closer to its expected level of poverty predicted by the EPRP. Thus, if we interpret the distances marked in Figure \ref{fig:hc_vs_EPRP_high} as an approximation to the speed of convergence of an economy to its predicted poverty levels, we would expect Colombia to reduce its poverty more rapidly than Costa Rica. This intuition is confirmed by Figure \ref{fig:headcount_three_countries}, where we notice that Colombia  decreased its poverty levels in the period 2010 to 2018, approaching the values of Costa Rica; whereas, Costa Rica did not change its poverty levels in a significant manner in the same period. Nonetheless, the difference of poverty levels in 2018 and the difference of their gaps concerning their expected poverty levels (Figure \ref{fig:hc_vs_EPRP_high}) let us conclude that Colombia has potential (thanks to its productive structure) to continue reducing its poverty levels, at least up to the same levels of Costa Rica.\\

On the other hand, despite China seems to have a lesser value of EPRP than Colombia, let us recall that this regression was performed with an average of the EPRP. Therefore, China started with lesser possibilities in a moment, but then it reached much better possibilities than the other countries. With an average value of its EPRP, the prediction suggests that China should continue improving its poverty levels in the future. It is reflected in Figure \ref{fig:headcount_three_countries}, where in 2010 China had higher poverty values than Colombia and Costa Rica, but during 2010 to 2018, it reduced more rapidly their poverty levels situating it in a better position than Colombia and Costa Rica. 

\section{Discussion and conclusions}

In this work, we studied poverty traps from the perspective of production. We proposed a method to project monetary poverty to products inspired by \cite{hartmann2017linking}. To do this, we combined ideas from complexity and relatedness such as the Product Space (PS) (\citealp{hidalgo2007product}; \citealp{hausmann2011network}) to create measures that capture to what extent the monetary poverty of a country is explained by its productive structure.  We introduced two types of measures at product level: the product poverty index (PPI) that captures short-run patterns of poverty and the Eigenpoverty index, that exploits the entire structure of the PS to predict long-run poverty patterns. We found through regression analysis that these two measures are strong predictors of  present and future levels of poverty in countries. Our work distinguishes from previous works in two main points. First, we understand poverty traps in terms of poverty measures instead of as stagnation of the GDP. And second, we study poverty traps from a nonaggregative point of view, specifically using the PS to show how different (and nuanced) compositions of the exports basket of a country can affect the dynamics of its monetary poverty levels.  \\

We depicted our findings through three illustrative examples (Colombia, Costa Rica, and China), where we showed how our measures can be used and how they capture nuances of poverty in terms of the export basket of countries. For this aim,  we considered two measures for countries based on the PPI (short-run measure) and the Eigenpoverty index (long-run measure). We considered the comparison between Colombia and Costa Rica. The Colombian economy has been historically more diverse than the Costa Rican economy, although simultaneously with higher poverty levels. In terms of our measures, we found that their potential to reduce poverty in the short-run has fluctuated significantly through time. On the other hand, their long-run reduction potential has tended to converge in the last few years, with Colombia still exhibiting a higher potential to reduce its poverty in the long term, but, at the same time, presenting higher poverty levels than Costa Rica. These patterns are coherent with their poverty (headcount measure) behavior through time since Colombian poverty has decreased consistently approaching , in average, Costa Rica levels. Our findings are consistent with what has been found in well-documented studies that have shown the negative impact of the production of goods of extractive nature in the poverty poverty levels of countries (\citealp{acemoglu2001colonial}; \citealp{sachs2001curse}; \citealp{apergis2018poverty}). \\

Furthermore, thanks to our non aggregative approach, our methodology allows us to consider more nuanced cases like the case of Colombia and Costa Rica discussed above. In this case, neither of these economies are in a dramatic situation (e.g., doomed by their natural resources endowments) and are significantly diverse with an amount of exports with \textbf{RCA>1} of 144 goods for Colombia and 130 goods for Costa Rica in 2010. The subtlety here relies on the nature of the composition of the export basket of Colombia that favors the production of agricultural and mineral products, textiles associated with natural resources, chemicals, and plastics. In contrast, Costa Rican economy has specialized some agricultural products, foodstuffs, and chemicals. In cases of this sort, the difference in their poverty levels lies in the type of products they produce, rather than their diversity.\\

We have also contributed to the theoretical literature in relatedness theory introducing long-run measures that exploit the whole vicinity structure of the PS. Commonly, the measures in complexity and relatedness theory to capture the influence of a product or the diversification opportunities of a country, have been constructed only considering first order neighbors. Examples of these measures are the density ratios or centrality metrics proposed in \cite{hidalgo2007product}, the complexity outlook measure \citep{hausmann2014atlas}, regional cohesion measures \citep{neffke2011regions} and others (for a detailed account of other measures see, e.g.,  \citealp{hidalgo2021economic}). In this sense, our measure is closer to the complexity or fitness measures that consider the entire network structure to assign values to products independently of the country under study.\\

Although this work provides elements to detect poverty based on the basket exports of countries,  it does not give concrete paths of diversification to mitigate poverty. Nonetheless, our conclusions open the opportunity to address the challenging problem of poverty reduction from the perspective of diversification. A feasible strategy to fight poverty in countries with a significantly diversified export basket could consist of identifying sets of target goods with low poverty levels to gain, in the terminology of this work, poverty reduction potential. This amounts to considering applications of our findings in the context of optimal diversification \cite{alshamsi2018optimal} or guided diversification \cite{church2020p} to help countries to improve their poverty levels through selective innovation. Those methods combined with the metrics developed in this work for product poverty can provide a guideline for countries to reduce their poverty levels and help them to avoid falling into a poverty trap without neglecting other objectives in their industries.

\section{Appendix}


\subsection{Time interval} \label{sec:timeinterval}

\begin{figure}[h!]
    \centering
    \includegraphics[trim={0cm 1cm 0cm 1cm},clip,scale=0.45]{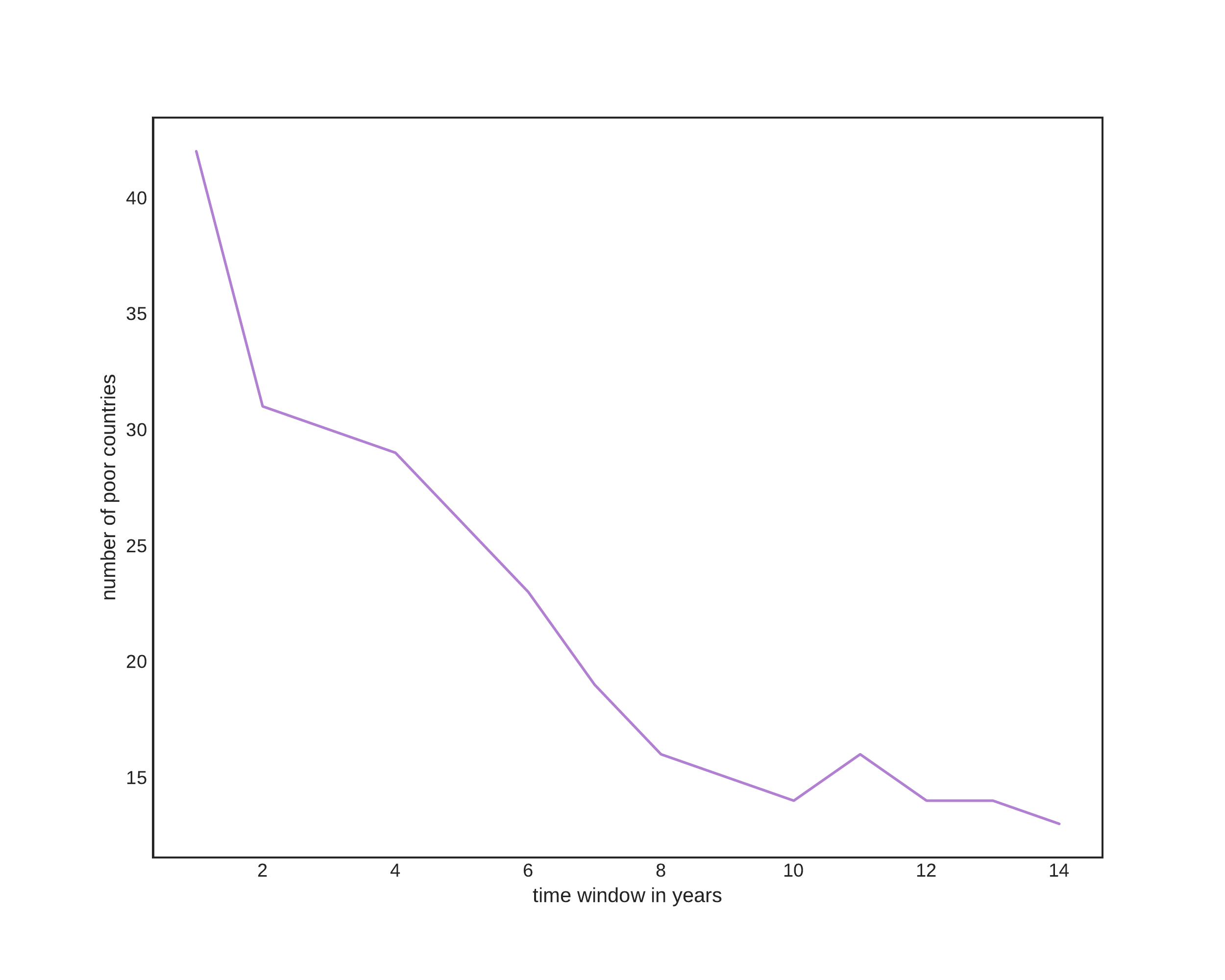}
    \caption{Elbow of the number of countries that remained in high levels of poverty using different time intervals.}
    \label{fig: elbow}
\end{figure}

We used eight years as the time interval to analyze the stagnation of poverty based on an empirical analysis of the poorest countries. First, we computed the average headcount index between 1981 and 2018 for 166 countries. Then, we filtered those economies with an average poverty ratio greater than 50\%. The latter restricts the analysis to 79 economies. Next, we analyzed for different time intervals how many countries only varied their poverty more than -3\% (capturing countries with minor variations or that even increased their poverty rates). Also, this threshold was chosen based on the average percent changes in poverty of the 79 countries year to year. Figure \ref{fig: elbow} illustrates for different time windows how many poor countries had not changed or increased their poverty levels. Thus, since this graph stabilizes after eight periods, we chose this number as our time frame. 

\subsection{Poverty measures} \label{sec:povertymeasures}

This section is devoted to explain why we chose the headcount measure as our poverty variable for studying poverty traps and briefly reviewing poverty indices. \\

We recognize that poverty is a multidimensional phenomenon. Therefore, it is ideal to consider multidimensional measures and monetary measures to compare which of these would be more precise for the type of analysis that we aimed to develop. However, data availability plays an important essential role in this decision. Usually, poverty was thought to be a phenomenon that needed to be monitored only in developing countries. The latter directly affects data availability in public repositories for studying poverty worldwide. Recent recommendations of Prof. Atkinson's commission on Global Poverty \citep{atkinson2017monitoring} to the World bank  appeal to measure poverty worldwide. In response, the World Bank developed a tool called PovcalNet in which some poverty measures are available for 166 countries including 28 high-income countries, from 1967 until 2018. These poverty indices include only monetary measures (Headcount, Gap, Gap Squared and Watts indices). The monetary measures consider the poverty line. It is established based on the minimum income needed to purchase a set of essential goods: food, clothes, shelter, water, electricity, education, and reliable healthcare. Thus, it implicitly takes into account some non monetary aspects.\\

 The following review is based on \cite{haughton2009handbook} and \cite{morduch2006concepts}. There are several indicators of two main streams: monetary and nonmonetary measures. Nonmonetary measures, such as MPI (Multidimensional Poverty Index) and HPI (Human Poverty Index), seek to go beyond the lack of income or basic consumption considering several dimensions of poverty such as health, mortality ratios (before 40 years), years of schooling, housing conditions, etc. However, there is not data available for non-poor economies, and since these measures considers several variables, some countries lack for some of them making difficult a cross-country comparability. On the other hand, monetary measures are indices based on consumption or income. Examples of these latter measures (already mentioned) include the headcount index (also called poverty incidence or poverty rate), the poverty gap index, the Watts index, the Foster-Greer-Thorbecke poverty index, and the poverty squared gap index. \\
 
 Monetary measures need to satisfy some axioms or properties to assure a proper measurement of the phenomena. The first property alludes to scale invariance, in which if the population rise and everything else remains constant, then the poverty measure should stay unchanged. The second property is the focus axiom, in which changes in the better-off people should not affect the poor population. Thus, measures with relative poverty lines are overlooked. The third is the monotonicity axiom which states that if the poor people's income decrease, then the measure should increase. The fourth is the transfer axiom, which refers to the transference of income among the poor, in which if a poor person transfers income to a poorer person, the measure should decrease. And lastly, the decomposition axiom states that the measure should allow to decompose poverty by subpopulation. The fulfilment of these properties guarantees a minimum level of consistency for monetary measures which, in conjunction with  data availability and homogeneity in the measurement among countries, turns them into a suitable tool to compare poverty at international level.\\

 The headcount ratio measures the proportion of people below the poverty line as

$$ P_0 =  \frac{1}{N} \sum_{i=1}^{N}I(y_i<z) = \frac{N_p}{N}$$ 

 where $N$ is the total population, $z$ is the poverty line, $y_i$ is the welfare indicator,i.e., consumption per capita or income, $I(.)$ takes the value of $1$ when income, consumption falls below the poverty line and $0$ otherwise; and $N_p$ is the number of poor people. The headcount ratio is the simplest to compute and is also a powerful descriptive tool of poverty. However, it fails the transfer axiom, and also it may give incentives to policy makers to implement strategies in the eradication of poverty for those that are below the poverty line, but close to surpass it. \\
 
On the other hand, the gap Index measures by how much income an individual is below the poverty line. Thus, it measures the depth of poverty, although this interpretation is given when the shortfall is divided by the poor population. Commonly the poverty gap is calculated as 

$$P_1 = \frac{1}{N}\sum_{i=1}^{N}  \left(\frac{y_i-z}{z}\right)I(z-y_i).$$

Let us notice that this measure is not simple to interpret and also violates the transfer axiom, but it can captures improvements from the most deprived people. The violation of the transfer axiom can be sorted raising the individual gaps to a power greater than 1 as

$$P_\alpha = \frac{1}{N}\sum_{i=1}^{N}  \left(\frac{y_i-z}{z}\right)^\alpha I(z-y_i).$$

This measure is known as the Foster-Greer-Thorbecke poverty index, which provides sensitivity according to the $\alpha$ chosen. It is common to find $\alpha=2$, and this is the poverty squared gap index. The Watts index also considers the transfer axiom, and it is computed as $$ W =  \frac{1}{N}\sum_{i=1}^{N} [\ln(z)-ln(y_i)]I(z-y_i).$$

Although the Foster-Greer-Thorbecke poverty index, the poverty squared gap index, and the Watts index are sensitive measures, they lack for easy interpretability. For this reason and the lack of data in multidimensional indices, we chose the headcount measure as our frame index.\\

\newpage
\bibliographystyle{apalike}
\bibliography{bibliography.bib}

\end{document}